\documentclass[10pt,journal,compsoc]{IEEEtran}

\ifCLASSOPTIONcompsoc
  \usepackage[nocompress]{cite}
\else
  \usepackage{cite}
\fi
\usepackage{cite}
\usepackage{amsmath,amssymb,amsfonts}
\allowdisplaybreaks
\usepackage{algorithm, algorithmic}
\usepackage{graphicx}
\usepackage{textcomp}
\usepackage{xcolor}
\usepackage{optidef}
\usepackage{subcaption}
\usepackage{lipsum,graphicx}
\usepackage[font={small}]{caption}
\usepackage[english]{babel}
\usepackage{environ}         
\usepackage{etoolbox}        
\usepackage{xparse}
\usepackage{etoolbox}
\usepackage{xcolor}

\newtheorem{definition}{Definition}[section]

\ifCLASSINFOpdf
\else
\fi
\begin{document}
%
\title{\huge Satellite-based ITS Data Offloading \& Computation in 6G Networks: A Cooperative Multi-Agent Proximal Policy Optimization DRL with Attention Approach}
%
%
%
%

\author{Sheikh Salman Hassan,~\IEEEmembership{Student Member,~IEEE,} Yu Min Park, 
Yan Kyaw Tun~\IEEEmembership{Member,~IEEE,} Walid Saad~\IEEEmembership{Fellow,~IEEE,} Zhu Han~\IEEEmembership{Fellow,~IEEE,}
and~Choong Seon Hong,~\IEEEmembership{Senior Member,~IEEE,}
\IEEEcompsocitemizethanks{\IEEEcompsocthanksitem Sheikh Salman Hassan, Yu Min Park, and Choong Seon Hong are with the Department of Computer Science and Engineering, Kyung Hee University, Yongin, 446-701, Republic of Korea.\protect\\
E-mail: \{salman0335, yumin0906, cshong\}@khu.ac.kr
\IEEEcompsocthanksitem Yan Kyaw Tun is with the School of Electrical Engineering and Computer Science, KTH Royal Institute of Technology, Stockholm, Sweden. \protect\\
E-mail:  yktun@kth.se
\IEEEcompsocthanksitem Walid Saad is with the Department of Electrical and Computer Engineering, Virginia Tech, VA, 24061, USA.
E-mail:  walids@vt.edu
\IEEEcompsocthanksitem Zhu Han is with the Department of Electrical and Computer Engineering, University of Houston, Houston, TX 77004-4005, USA. \protect\\
E-mail:  zhan2@uh.edu

}
}

%
%

\markboth{Journal of \LaTeX\ Class Files,~Vol.~14, No.~8, August~2015}%
{Shell \MakeLowercase{\textit{et al.}}: Bare Advanced Demo of IEEEtran.cls for IEEE Computer Society Journals}
%



\IEEEtitleabstractindextext{%
\begin{abstract}
The proliferation of intelligent transportation systems (ITS) has led to increasing demand for diverse network applications. However, conventional terrestrial access networks (TANs) are inadequate in accommodating various applications for remote ITS nodes, i.e., airplanes and ships. In contrast, satellite access networks (SANs) offer supplementary support for TANs, in terms of coverage flexibility and availability. In this study, we propose a novel approach to ITS data offloading and computation services based on SANs. We use low-Earth orbit (LEO) and cube satellites (CubeSats) as independent mobile edge computing (MEC) servers that schedule the processing of data generated by ITS nodes. To optimize offloading task selection, computing, and bandwidth resource allocation for different satellite servers, we formulate a joint delay and rental price minimization problem that is mixed-integer non-linear programming (MINLP) and NP-hard. We propose a cooperative multi-agent proximal policy optimization (Co-MAPPO) deep reinforcement learning (DRL) approach with an attention mechanism to deal with intelligent offloading decisions. We also decompose the remaining subproblem into three independent subproblems for resource allocation and use convex optimization techniques to obtain their optimal closed-form analytical solutions. We conduct extensive simulations and compare our proposed approach to baselines, resulting in performance improvements of $9.9\%$, $5.2\%$, and $4.2\%$, respectively. 
\end{abstract}

}

\maketitle
\IEEEdisplaynontitleabstractindextext
\IEEEpeerreviewmaketitle
\ifCLASSOPTIONcompsoc
\IEEEraisesectionheading{\section{Introduction}\label{sec:introduction}}
\else
\section{Introduction}
\label{sec:introduction}
\fi
\IEEEPARstart{T}{he} terrestrial access networks (TANs) are utilized in developed regions with high population densities for economic benefits. But TANs are unable to cover the wide airspace, sea, and desert regions due to network infrastructure unavailability. When it comes to the \textit{connected world} objective of the sixth generation (6G) networks  \cite{6G_PROF_WALID}, coverage becomes more important than intensity \cite{why_satellite}. The 6G networks will provide applications to diverse sectors, e.g., transportation, industrial, and energy. Extending connection to the rest of the zones has become critical to going forward with future networks. Recently, research has been conducted on integrated space-terrestrial networks (ISTNs) \cite{ISTN1,ISTN2}. Low Earth orbit (LEO) \cite{LEO_sat1} and cube satellites (CubeSats) \cite{maral2020satellite} are the most practical configurations for networks' applications since they orbit closer to the Earth and offer low latency \cite{LEO_latency}.

Mobile edge computing (MEC) could be a successful paradigm for real-time intelligent transportation systems (ITSs) by delivering communication, computing, and caching resources from nearby satellites \cite{MEC_replace,why_satellite2}. However, data-driven ITS services are delay-sensitive and computing-intensive. Thus, mobile ITS nodes (e.g., ships, trains, airplanes, and vehicles) need efficient data communication and computing at MEC-enabled satellite servers. Although these servers are placed close to the ITS nodes, a tight need for communication and computing resource allocation exists \cite{MEC_survey}. Similarly, ITS nodes can experience service delays due to unanticipated network congestion, unequal load balancing, dynamic network architecture, and uneven distribution.

Recently, the development of offloading techniques for computing data in MEC-enabled networks has received significant attention \cite{related_mec_1,related_mec_2,related_mec_ris_uav_1}. Various MEC-enabled services are proposed in the literature, i.e., aerial-MEC \cite{related_mec_uav_1} and vehicular-MEC \cite{related_mec_vehicular_1}. Moreover, a few works investigated the MEC-enabled satellite networks, as given in \cite{related_sat_mec_1, related_sat_mec_2}. These works suggested MEC-based offloading techniques \cite{mec_offloading}, i.e., the multi-user game model \cite{related_MEC_V2I} and reinforcement learning \cite{related_MEC_RL}, to increase dependability. But these studies overlooked the contest for MEC resources among several tasks and simply optimized offloading decisions. Moreover, literature on MEC-enabled satellite networks by considering different orbiting satellites with various computing and communication resources is also missing \cite{sat_mec_cost,sat_mec_signal}.

To address the above-mentioned challenges, a few researchers developed resource allocation algorithms based on convex optimization \cite{MEC_Sat_convex_opt} and the semi-Markov decision process \cite{SMDP_SAT_MEC}. A few researchers have also built a combined optimization framework by incorporating task offloading with resource allocation \cite{sat_mec_resource_allocation}. Similarly, a few scheduling techniques, i.e., deep Q-learning \cite{mec_sat_q_learning} and alternating direction method of multipliers (ADMM) \cite{sat_mec_ADMM} are also studied. However, such integrated optimization frameworks are based on NP-hard non-linear programming, which cannot be solved efficiently in polynomial time \cite{Own_NOMS_NP}. Moreover, these studies are based on a centralized controller, which is impractical due to excessive communication costs and scheduling complexity for large-scale networks.

This research examines a unique service scenario of data-driven ITS task offloading to MEC-enabled satellite networks, in which data sensed by ITS nodes is offloaded to diverse computing servers, i.e, Geosynchronous-orbit (GEOs), LEOs, and CubeSats, motivated by the aforementioned challenges. The size of the ITS node's data collection and the needed computing resources distinguish data-driven tasks, which are separated into various sub-tasks based on the statistical distribution across nodes. To simulate the task offloading procedure, three different types of communication and computing models are considered with the heterogeneity of computing servers. The communication (bandwidth) and computing (CPU) resource rental prices, which differ depending on the type of computing server, are also taken into account. To minimize service time and the task computing price concurrently by fully utilizing diverse communication and computing resources, LEO and CubeSats are responsible for making distributed scheduling for surrounding ITS nodes inside their coverage, including offloading server and resource allocation.

The challenges mentioned above need to be addressed effectively. One major challenge is achieving a balance between two opposing goals, as completing a task quickly requires paying a higher price for additional computing resources. Additionally, satellites must regularly interact with each other and with ITS nodes to retain current global information, which can result in excessive communication prices. Another challenge is optimizing scheduling, which may require a higher temporal complexity, especially in large networks, by combining task offloading with resource allocation. To address these challenges, we propose a distributed task offloading algorithm based on cooperative multi-agent proximal policy optimization (Co-MAPPO) deep reinforcement learning (DRL) \cite{ISTN_RL1} with an attention approach. The attention approach encodes various observations and is designed by the network operator to provide each agent with a differentiated fit reward. Furthermore, we use decomposition and convex theory to construct the optimal resource allocation solution, which is distributedly implemented on each satellite. The proposed approach has several advantages. Firstly, it balances the trade-off between completion time and price by efficiently allocating computing resources. Secondly, it facilitates inter-satellite communication and information sharing while minimizing communication costs. Thirdly, it optimizes scheduling by combining task offloading with resource allocation, ensuring efficient use of resources. Overall, this approach can improve the performance and scalability of multi-layer satellite-based communication systems.
The following are the main contributions of this paper:

\begin{itemize}
    \item In this paper, we propose a service architecture for data-driven ITS task offloading and computation to MEC-enabled diverse and multi-layer satellite access networks (SANs).
    \item On this basis, we formulate a joint delay and rental price minimization problem for different satellite servers while optimizing offloading task selection, computing, and bandwidth resource allocation.
    \item To handle the formulated mixed-integer non-linear programming (MINLP) problem, which is NP-hard, we propose a two-stage algorithm based on the Co-MAPPO DRL algorithm in cooperation with the attention approach and convex theory.
    \item Extensive simulations are used to demonstrate the efficiency and effectiveness of our proposed methodology in achieving the desired objective function when compared to baseline approaches.
\end{itemize}

The rest of this paper is organized as follows. Section \ref{rel_work} studied the current literature on ITS and satellite-based data offloading, processing and summarized their key limitations. Section \ref{pre_sys} describes the proposed system model and the formulated optimization problem is given in Section \ref{prob_form}. Section \ref{prop_algo} introduces the proposed algorithm to solve the formulated problem. The performance evaluation of the proposed algorithm is presented in Section \ref{sim_result}, and we conclude our paper in Section \ref{conc}.  The abbreviations and notations used in this work are summarized in Table \ref{acronyms} and \ref{notations}, respectively.

\section{Related Work}
\label{rel_work}
\subsection{Data Offloading in ITS Wireless Networks}
To address the issue of ITS data offloading, several studies have examined various algorithms, including heuristics and learning-based approaches, for application in terrestrial wireless networks. The authors in \cite{new_rel_1_offRefBlue} proposed an intelligent software-defined (SD) cellular vehicle-to-everything (C-V2X) network framework to relieve congestion and improve load balance, enabling flexible and low-complexity traffic offloading by decoupling the data plane from the control plane. Cellular and vehicle-assisted offloading are jointly performed in the data plane, while deep learning reduces SD control complexity and improves offloading efficiency in the control plane. The work in \cite{new_rel_2_offRefBlue} describes a data transmission network architecture based on the Manhattan mobility model, where data is transmitted between geographically distant data centers through vehicles on fixed routes. The authors propose using the temporal convolutional network (TCN) model to predict the weight of delay allocation and a genetic algorithm based on a reinforcement learning mechanism (RLGA) to pre-allocate resources for offloading requests.  The study in \cite{new_rel_3_offRefBlue} integrates three computing layers (vehicles, terrestrial network edges, and HAPS) to create an ITS computation framework, with the HAPS data library serving as the repository for essential application data. The work \cite{new_rel_4_offRefBlue} dealt with vehicular edge computing (VEC) by jointly selecting a network and offloading computation for minimizing the overall latency and energy consumption while considering energy-saving mechanisms on both the user and infrastructure sides. The work \cite{new_rel_5_offRefBlue} introduces the predicted k-hop-limited Multi-RSU-Considered (PKMR) approach for Vehicle-to-Vehicle to Roadside Unit (RSU) data offloading, utilizing the Software Defined Network (SDN) controller architecture integrated within the Multi-Access Edge Computing (MEC) server.

\subsection{Satellite-based Wireless Access Networks}
Numerous heuristic algorithms have been developed to enhance the management of resource allocation within SANs. The authors in \cite{Sat-MEC1} suggested satellite MEC (SMEC), which allows the user equipment to get benefit from MEC services across satellite communications when ground MEC is not available. The authors in \cite{Sat-MEC2} introduced the radio access network's (RAN) cache-enabled  LEO satellite network. The authors suggested an integrated satellite-terrestrial cooperative transmission strategy to provide an energy-efficient RAN by offloading traffic from base stations (BSs) through satellite broadcast transmission. According to a concept put out by the authors in \cite{Sat-MEC3}, space-air-ground-integrated networks (SAGINs) are the future of edge computing because they effectively provide seamless coverage and effective resource management. The authors of \cite{Sat-MEC4} suggested a satellite-marine cloud-edge-terminal (CET) architecture that takes into account the portability of maritime vehicles while managing network resources. This architecture includes proactive caching, CET offloading, and integrated transmission optimization. This study gave marine network management a fresh viewpoint to better accommodate the movement of maritime communication entities. Additionally, in our recent studies \cite{Sat-MEC5, Salman_globecom}, we created space-air-sea (SAS) non-terrestrial networks (NTNs) for resource allocation that is energy-efficient. By cooperatively improving user equipment (UE) association, power control, and unmanned aerial vehicle (UAV) deployment, the aim is to optimize system energy efficiency (EE). In \cite{new_rel_1_satRefBlue}, authors studied a space-air-ground (SAG) integrated three-layer heterogeneous network model to maximize the sum rate of ground IoT devices, which further enhances the deep integration of communication and computation resources. Drawing from the literature presented above, heuristic algorithms necessitate a considerable undertaking in the development of heuristic rules or search strategies that are reliant on expert knowledge. This is primarily conducted concerning satellite-ground node association and resource allocation.

\subsection{DRL in Satellite Wireless Networks}
In recent years, scholars have employed DRL to automatically learn heuristics for solving combinatorial optimization problems by constructing MDPs and DRL policy networks, which serve as decision-making agents. The work \cite{new_rel_1_sat_DRLRefBlue} introduces the SAGINs framework that supports edge-computing Internet of Vehicles (EC-IoV) services in remote areas, intending to minimize task completion time and satellite resource usage. They reduce the action space with a pre-classification scheme and propose a deep imitation learning-driven offloading and caching algorithm for real-time decision-making. The study \cite{new_rel_2_sat_DRLRefBlue} presents a framework for resource allocation in terrestrial-satellite networks using non-orthogonal multiple access (NOMA). Additionally, a local cache pool deployment strategy is proposed to reduce time delay and enhance energy efficiency. To achieve this, the authors utilize a multi-agent deep deterministic policy gradient (MADDPG) method that incorporates user association, power control, and cache design for optimal energy efficiency. The study \cite{new_rel_3_sat_DRLRefBlue} examines that efficient power allocation between common and private streams is essential for improved rate-splitting multiple access (RSMA) performances in the LEO satellite, but is challenging due to limited and uncertain channel distribution information. To address this issue, a PPO-based scheme is proposed to optimize the sum rate of the system without any prior knowledge, resulting in an effective solution. The authors \cite{new_rel_4_sat_DRLRefBlue} propose a dynamic beam pattern and bandwidth allocation scheme using DRL for beam-hopping satellite systems. A cooperative multi-agent MADRL framework is presented, where agents are responsible for illumination or bandwidth allocation for a single beam. The authors \cite{new_rel_5_sat_DRLRefBlue} suggest a dynamic system model that uses high altitude platforms (HAPs) with MEC servers and a backhaul system of LEO satellites to provide computation offloading and network access for V2V communications. Their goal is to minimize the total computation and communication overhead of the system by proposing a decentralized value-iteration-based reinforcement learning (RL) approach as a solution. The paper \cite{new_rel_6_sat_DRLRefBlue} focuses on optimizing limited satellite-based radio resources for efficient transmission and low complexity. To handle the complexity of the problem, the single-agent deep reinforcement learning approach is extended to a cooperative multi-agent deep reinforcement learning method. Finally, the authors \cite{new_rel_7_sat_DRLRefBlue} propose a Multi-Agent Inter-plane Inter-satellite Links Planning (MA-IILP) method that utilizes Multi-Agent Deep Deterministic Policy Gradient (MADDPG) algorithm to optimize total throughput and inter-plane ISL switching rate. The primary obstacle in utilizing DRL within multi-layer SANs is the development of an MDP capable of handling intricate constraints and the creation of an efficient policy network with satisfactory generalization capabilities. Thus far, no efforts have been made to address the challenges associated with the application of Co-MAPPO in ITS.
\begin{table}[t]
	\centering
	\caption{Summary of Abbreviations.}
	\label{acronyms}
        \begin{tabular}{p{0.15\columnwidth}p{0.75\columnwidth}}
		\hline 
		Abbreviation & Meaning\\
		\hline \hline
		ITS & Intelligent Transportation System\\
        LEO & Low-Earth Orbit\\
        GEO & Geosynchronous-Earth Orbit\\
        MEC & Mobile Edge Computing\\
        MINLP & Mixed-Integer Non-Linear Programming\\
        Co-MAPPO & Cooperative Multi-Agent Proximal Policy Optimization\\
        DRL & Deep Reinforcement Learning\\
        TAN & Terrestrial Access Network\\
        SAN & Satellite Access Network\\
        6G & Sixth Generation\\
		ISTN & Integrated Space-Terrestrial Network \\
        NP-Hard & Non-Deterministic Polynomial Time \\
        CTE & Crowed-sourced Transportation Entity \\
        CubeSat & Cube Satellite \\
        LMS & LEO satellite-based MEC Server \\
        CNS & Core network Server (based on GEO satellite layer) \\
        MST & Mean Service Time \\
        MSP & Mean Service Price \\
        KKT & Karush–Kuhn–Tucker (conditions) \\ 
		\hline
	\end{tabular} 
 \vspace{-0.1in}
\end{table} 

In the context of multi-layer satellite resource allocation, collaboration among satellites (agents) is a crucial consideration. However, heuristic algorithms have a slow convergence time and are unable to achieve real-time decisions. Previous studies have not simultaneously examined offloading of ITS data and computation on multi-layer satellites while considering delay and price requirements. Additionally, the Co-MAPPO DRL technique, which is highly effective for learning stochastic policy in complicated and dynamic satellite environments, has not been utilized in the literature. To address these gaps, we propose a collaborative data management framework for diverse satellite network environments. Specifically, we introduce an attention mechanism based on the Co-MAPPO DRL model to solve the ITS data management problem. Our approach obviates the need for designing heuristics that rely on expert knowledge, instead leveraging a large number of learning instances to enhance the models. After training, the models learn a policy for solving the proposed problem, which has the potential to generalize to other ITS problems with different characteristic distributions without requiring additional learning or parameter adjustments.

\section{System Model and its Preliminaries}
\label{pre_sys}
\begin{figure}[t]
    \centering
    \includegraphics[width=\columnwidth]{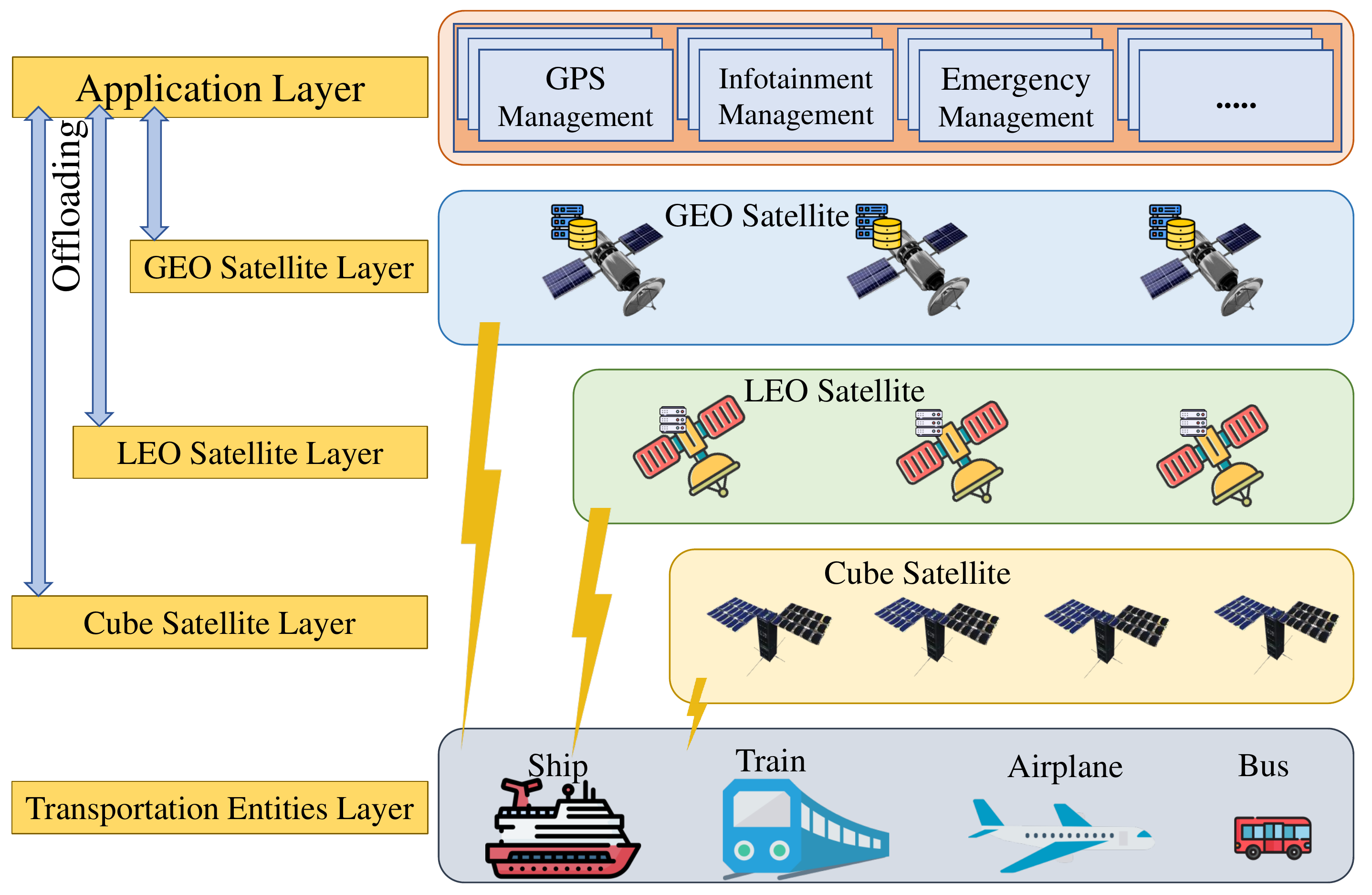}
    \caption{Illustration of the service architecture enabling data-driven ITS task offloading to MEC-enabled satellite networks.}
    \label{sysmod}
    \vspace{-0.2in}
\end{figure}

We presented our service architecture enabling data-driven ITS task offloading to MEC-enabled satellite networks in Fig. \ref{sysmod}. The proposed system framework is composed of an application layer, transportation entities layer, CubeSats layer, LEO satellite layer, and GEO satellite layer, which are defined below in the respective order:

\textbf{Application Layer:} This layer handles a range of ITS functions, e.g., traffic light control, abnormal traffic detection, weather updates, global positioning system (GPS) updates, emergency notifications, and traffic flow forecasts to name a few. These are data-driven tasks that require the processing of transportation data acquired by transportation entities (ITS nodes) deployed across each transportation path network, i.e., maritime, terrestrial, aerial, underwater and space. Thus, a data-driven task is divided into multiple sub-tasks depending on data distribution. Each sub-task is linked to a collection of transportation data seen by ITS nodes and can be processed at the same time. The task can only be completed once all of the sub-tasks have been processed. The MEC-enabled satellite services architecture has the advantage of offloading sub-tasks to neighboring satellites without centralized processing at the core network (i.e., the GEO satellite\footnote{In this network architecture, we consider GEO satellite as network controller due to its more powerful network resources and controlling which could easily handle LEO and Cube satellites under its coverage.}), significantly lowering communication time and enhancing computing efficiency. In this study, task communication and computing are used in the task offloading procedure. However, due to its small size, the computing result's recovery time is ignored \cite{recovery_time}.

\textbf{Transportation Entities Layer:} This layer is composed of crowd-sourced transportation entities (CTEs) i.e., ships, trains, aircraft, and vehicles. These CTEs can collect various kinds of transportation data since they are equipped with a range of smart sensors. Data for numerous sub-tasks that need to be wirelessly offloaded to nearby computing satellite servers first can be stored in a CTEs.

\textbf{Cube Satellite Layer:} This layer is composed of MEC-enabled CubeSats, which serve two purposes: as computing servers and as local schedulers. CubeSats are required to be equipped with a processor and be able to do the local computing for sub-tasks within the communication range of CTEs. However, only a small number of sub-tasks can be offloaded to a CubeSat at a time because the operational design (miniature form factor) of CubeSats mandates a low orbiting height, necessitated by their limited functionalities and onboard processing capabilities. Furthermore, the communication bandwidth available to CubeSats is restricted due to the relatively low onboard processing capabilities that can accommodate only a minimal number of CTE requests and processes \cite{cubesats_XRefBlue}. Therefore, CubeSat rental prices are considered the most affordable.

\begin{table}[t]
	\centering
	\caption{Summary of Notations.}
	\label{notations}
        \begin{tabular}{p{0.1\columnwidth}p{0.8\columnwidth}}
		\hline 
		Notation & Definition\\
		\hline \hline
		$\mathcal{H}$ & Set of CNS,  $|\mathcal{H}|= H$\\
		$\mathcal{L}$ & Set of LMS,  $|\mathcal{L}|= L$\\
		$\mathcal{C}$ & Set of CubeSats,  $|\mathcal{C}|= C$\\
        $\mathcal{E}$ & Set of CTEs,  $|\mathcal{E}|= E$\\
        $\mathcal{C}_{\mu}$ & Set of CubeSats under LMS coverage,  $|\mathcal{C}_{\mu}|= C_{\mu}$\\
        $\mathcal{E}_{\mu}$ & Set of CTEs under LMS coverage,  $|\mathcal{E}_{\mu}|= E_{\mu}$\\
		$\mathcal{D}$ & Set of Data-driven tasks, $|\mathcal{D}|= D$\\
		$d \in \mathcal{D}$ & Single task from set of data-driven tasks $\mathcal{D}$\\
		$M_d$ & Data memory requirement to execute task $d$ \\
		$\nu_d$ & Computing power requirement to execute task $d$ \\
		$d_{\mu} \in d$ & Task $d$ division into $d_{\mu}$ sub-tasks by following the distribution of cached memory among various CTEs \\
		$M_{d_{\mu}}$ & Assigned data memory stored by CTE e for $d_{\mu}$ \\
        $\nu_{d_{\mu}}$ & Desired computing power for $d_{\mu}$ \\
        $\mathcal{B}_{\mu}$ & Set of all accessible satellites as $\mathcal{B}_{\mu} = \{\mathcal{L} \cup \mathcal{C} \cup \mathcal{H}\}.$\\
        $x^b_{d_{\mu}}$ & Offloading variable decides satellite for computation $\forall b\in\mathcal{B}_{\mu}$ \\
        $\lambda_b$ & Each satellite $b$ computing power \\
        $\zeta_b$ & Each satellite $b$ wireless connection bandwidth \\
        $y_{d_{\mu}}^b$ & Fraction of CTE wireless bandwidth designated to sub-task $d_{\mu}$ \\
        $T^{\mathrm{tran}}_{d_{\mu}, b}(n)$ & Transmission time between $\mu$ to $b$ at each time slot $n$ \\
        $p_{\mu}$ & Transmit power of CTE $\mu$ \\ 
        $g_{b, \mu}$ & Channel gain between satellite $b$ and $\mu$ \\
        $N_0$ & Additive white Gaussian Noise \\
        $\chi^{\mathrm{tran}}_c$ & Unit price of utilizing a Ka-band link of CubeSats\\
        $\chi^{\mathrm{tran}}_l$ & Unit price of utilizing a Ka-band link of LMS\\
        $\chi^{\mathrm{tran}}_h$ & Unit price of utilizing a Ka-band link of CNS\\
        $P^{\mathrm{tran}}_{d_{\mu}, b}$ & Communication price of offloading $d_{\mu}$ to Cubesats, $~\forall b \in \mathcal{C}$ \\
        $P^{\mathrm{tran}}_{d_{\mu}, b}$ & Communication price of offloading $d_{\mu}$ to LMS $~\forall b \in \mathcal{L}$ \\
        $P^{\mathrm{tran}}_{d_{\mu}, b}$ & Communication price of offloading $d_{\mu}$ to CNS, $~\forall b \in \mathcal{H}$ \\
        $\omega_b$ & Computing resource of each processor, $b \in \mathcal{B}$ \\
        $\varrho_b$ & Number of processors per satellite server, $b \in \mathcal{B}$ \\
        $\beta^b$ & Percentage of computing resources allotted for sub-task $d_{\mu}$ \\
        $T^{\mathrm{comp}}_{d_{\mu}, b}$ & Computing time of sub-task $d_{\mu}$ at server $b$, $b \in \mathcal{B}$ \\
        $\nu_{d_{\mu}}$ & Required computing resources to execute $d_{\mu}$ at server $b$, $b \in \mathcal{B}$ \\
        $u_b$ & Required CPU cycles to compute $1$-bit of offloaded sub-task at server $b$, $b \in \mathcal{B}$ \\
        $T^{\mathrm{max}}_{d_{\mu}, b}$ &  Maximum allowable transmission and computation time for each task at server $b$ \\
        $\varrho_b$ & Maximum computing capacity of server $b$, $b \in \mathcal{B}$ \\
        $\chi^{\mathrm{comp}}_c$ & Unit rental computing price for CubeSat \\
        $\chi^{\mathrm{comp}}_l$ & Unit rental computing price for LMS \\
        $\chi^{\mathrm{comp}}_h$ & Unit rental computing price for CNS \\
        $P^{\mathrm{comp}}_{d_{\mu}, c}$ & Computing price of sub-task $d_{\mu}$ at CubeSat \\
        $P^{\mathrm{comp}}_{d_{\mu}, l}$ & Computing price of sub-task $d_{\mu}$ at LMS \\
        $P^{\mathrm{comp}}_{d_{\mu}, h}$ & Computing price of sub-task $d_{\mu}$ at CNS \\
        $T^{\mathrm{ser}}_{d_{\mu}}$ & Service time of each sub-task $d_{\mu}$ \\
        $P^{\mathrm{ser}}_{d_{\mu}}$ & Service price of each sub-task $d_{\mu}$ \\
        $T^{\mathrm{mean}}_{d}$ & Each task $d$'s service time as all the sub-tasks $d_{\mu} \in \mathcal{D}$ mean service time \\
        $P^{\mathrm{mean}}_{d}$ & Each task $d$'s service price as all the sub-tasks $d_{\mu} \in \mathcal{D}$ mean service price \\
        $\eta_1$ & Mean Service Time (MST) \\
        $\eta_2$ & Mean Service Price (MSP) \\
        $\eta$   & Weighted sum of MST and MSP (objective function)\\
		\hline
	\end{tabular} 
 \vspace{-0.1in}
\end{table}

\textbf{LEO Satellite Layer:} This layer is composed of LEO satellite-based MEC servers (LMSs) that fulfill two purposes: computing server and scheduler. Many pending sub-tasks that cannot be offloaded to CubeSats owing to limited computing power are offloaded to the LMS through a wireless CTE link and can be handled simultaneously. Due to the mobility of CTEs, the sub-task must be completed within the time limits of the CTE-LMS link. As a result, a large number of pending sub-tasks can compete for wireless bandwidth and computing resources. The LMS is anticipated to be more expensive to compute than CubeSats owing to its greater computing power. Furthermore, as the scheduler, the LMS is in charge of determining offloading choices for remaining sub-tasks, i.e., offloaded MEC satellite, communication bandwidth, and computing resource allocation, depending on sub-task information obtained by CTE reference signaling.

\textbf{GEO Satellite Layer:} This layer is composed of GEO satellites that function as core network servers (CNS) for the backbone network and considered with huge processing resources. If LMSs and CubeSats coverage are absent, the CTE can offload their sub-tasks to the CNS. The CTE can always contact the CNS owing to their global connection availability, but they must pay a higher communication price that is proportional to the amount of uploaded content. Furthermore, if the CTE fails to execute the associated sub-tasks within the LMS or CubeSats connection time, the CNS must be selected as the offloaded server. Similarly, the CNS is considered the most expensive computer server to rent.

According to the insights mentioned above, task completion involves coordination across various LMSs and CubeSats to integrate varied resources. Thus, a distributed technique for optimizing the use of various communication bandwidth and computing resources of CubeSats, LMSs, and CNSs must be designed and deployed on each MEC-enabled satellite to minimize both service time and price.
\subsection{ITS Network \& Data-driven Task Model}
In the considered system model, a task offloading in the satellite system follows the time division multiple access (TDMA), where the CNSs are represented as a set $\mathcal{H}$ of $H$ CNS, the LMSs are given as a set $\mathcal{L}$ of $L$ LMS, and the CubeSats are given as a set $\mathcal{C}$ of $C$ CubeSats, and the CTEs are represented as a set $\mathcal{E}$ of $E$ CTE in time duration $\mathcal{N}=\left\{1,2,..., N \right\} \in \mathbb{R}^{\vert N \vert}$. The number of CTEs and CubeSats under the coverage of LMSs $l$ is represented as $E_{\mu}$ and $C_{\mu}$, respectively. Thus, a set of CTEs and CubeSats neighboring to LMS $l$ is considered as $\mathcal{\mu}$, which is composed of $\mathcal{E}_{\mu}$ and $\mathcal{C}_{\mu}$, respectively. We consider a set that represents the data-driven tasks and can be given as $\mathcal{D}$, where each task $d \in \mathcal{D}$ is defined as a two-tuple $(M_d, \nu_d)$. Furthermore, $M_d$ and $\nu_d$ are data memory and computing power, respectively, to execute each task. Specifically, each given task $d$ is assumed to be further categorized into several sub-tasks by following the distribution of cached data memory $M_d$ between CTEs. In particular, each sub-task $d_{\mu}$ of $d$ is assigned a data memory $M_{d_{\mu}}$ stored by a CTE $e$. Therefore, the desired data memory $M_{d_{\mu}}$ is represented as $M_d \cap M_{\mu}$, and the desired computing power will be $\nu_{d_{\mu}}$, which is directly proportional to the size of $M_{d_{\mu}}$ (e.g., $\nu_{d_{\mu}} \propto \|M_{d_{\mu}}\|$). For the sake of generality, it has been considered that $\|d\|$ and $\|D\|$ represent the number of sub-tasks in $d$ and tasks in $\mathcal{D}$, respectively. 

\subsection{Data Driven ITS Task Offloading Decision Model}
Each sub-task $d_{\mu}$ offloading decision variable is denoted with notations given as a set of $x_{d_{\mu}}^b, ~\forall b\in \mathcal{B}_{\mu}$, where each $x_{d_{\mu}}^b$ represents whether offloaded $d_{\mu}$ is computing at satellite $b$ or not and $\mathcal{B}_{\mu}$ is the computing satellites set accessible for $\mu$, which can be defined as:
    $\mathcal{B}_{\mu} = \{b | b \in \mathcal{L} \cup \mathcal{C} \cup \mathcal{H}\}.$
We assumed that each sub-task could not be further subdivided and could only be allocated to one of the available computing satellites, which may be described as follows:
\begin{equation}
    x_{d_{\mu}}^b \in \{0,1\},\forall b\in\mathcal{B}_{\mu},~\sum_{\forall b\in\mathcal{B}_{\mu}} x_{d_{\mu}}^b = 1,\forall d \in \mathcal{D},\forall \mu \in \ \mathcal{E}_{\mu}.
\end{equation}
Moreover, each satellite $b$ has computing power $\lambda_b$ and wireless connection bandwidth $\zeta_b$ which can be defined as a two-tuple $(\lambda_b, \zeta_b)$. The CTEs $\mu~\in~\mathcal{E}_{\mu}$ are considered with cached data memory for task defined as $M_{\mu}$ and each CubeSat $\mu~\in~\mathcal{C}_{\mu}$ has been considered with computing power $\lambda_{\mu}$.  

\subsection{Data-driven ITS Task Communication \& Price Model}
In this part, we develop ITS sub-task communication models for LMS, CubeSats, and CNS. The wireless CTE's bandwidth (i.e., Ka-band) of each service-providing satellite (e.g., CNS, LMS, and CubeSats) has competed among multiple CTEs for task communication. When the sub-task $d_{\mu}$ is offloaded to any satellite $b$ (e.g., $x_{d_{\mu}}^b = 1$), then let's consider $y_{d_{\mu}}^b$ the ratio of CTE wireless bandwidth designated to sub-task $d_{\mu}$, where the total sum of all designated ratios must be within the threshold of one at each time slot $n$, which can be characterized as below:
\begin{equation}
    \sum_{\forall d \in \mathcal{D}} \sum_{\forall \mu \in \mathcal{E}_l} x_{d_{\mu}}^b(n) y_{d_{\mu}}^b(n) \leq 1, ~\forall b \in \mathcal{B}, ~\forall n \in \mathcal{N}.
\end{equation}
By providing the CTE and respective satellite wireless communication bandwidth as $\zeta_b$, the offloading sub-task $d_{\mu}$ time for communication between $\mu$ to $b$ at each time slot $n$ can be calculated as below:
\begin{equation}
    T^{\mathrm{tran}}_{d_{\mu}, b}(n) = \frac{\| M_{d_{\mu}}\|}{y_{d_{\mu}}^b(n)  \zeta_b  \log_{2} \big( 1 + \frac{p_{\mu}(n) g_{\mu, b}(n)}{N_0} \big)},
\end{equation}
where $p_{\mu}$ refers to the transmit power of CTE $\mu$, $g_{b, \mu}$ represents the channel gain from satellite $b$ to $\mu$, and $N_0$ is the additive white Gaussian Noise.

As a result, the communication price is as follows. Let $\chi^{\mathrm{tran}}_c$, $\chi^{\mathrm{tran}}_l$, and $\chi^{\mathrm{tran}}_h$ signify the unit price of utilizing a Ka-band link of CubeSats, LMS, and CNS, respectively, and then compute the communication price of offloading $d_{\mu}$ to each respective satellite can be characterized as the product of unit price and bandwidth allocation:
\begin{equation}
\begin{cases}
    P^{\mathrm{tran}}_{d_{\mu}, b} = \chi^{\mathrm{tran}}_c  y_{d_{\mu}}^b(n)  \zeta_b,~\forall b \in \mathcal{C} & \mathrm{CubeSat~ price},\\
    P^{\mathrm{tran}}_{d_{\mu}, b} = \chi^{\mathrm{tran}}_l  y_{d_{\mu}}^b(n)  \zeta_b,~\forall b \in \mathcal{L} & \mathrm{LMS~ price},\\
    P^{\mathrm{tran}}_{d_{\mu}, b} = \chi^{\mathrm{tran}}_h  y_{d_{\mu}}^b(n)  \zeta_b,~\forall b \in \mathcal{H} & \mathrm{CNS~price}.
\end{cases}
\end{equation}
where the unit for communication price is considered as cyc/sec/\$.

\subsection{Data-driven ITS Task Computing \& Price Model}
In this part, we develop ITS sub-task computing models for LMSs, CTEs, and CNSs. The time of a computing sub-task and the price of renting computing resources are discussed in detail. We assume that the LMS's and CNS's computing resources are divided across several pending sub-tasks, but CubeSats execute fewer sub-tasks at each time slot $n$ due to computing resource constraints. By providing the maximum computing resource of each processor (i.e., cycles/s) owned by each satellite's server $b$, which can be indicated as $\omega_b$. We consider that the number of processors per satellite server is $\varrho_b$, and the percentage of computing resources allotted for sub-task $d_{\mu}$ is $\beta^b$. Thus, the completion time of computing sub-task $d_{\mu}$ can be expressed as below:
\begin{equation}
    T^{\mathrm{comp}}_{d_{\mu}, b} = \frac{\nu_{d_{\mu}}}{\beta_{d_{\mu}}^b  \omega_{b}} = \frac{ u_b  M_{d_{\mu}}}{\beta_{d_{\mu}}^b  \omega_{b}},
\end{equation}
where $\nu_{d_{\mu}}$ is the required computing resources to execute $d_{\mu}$ sub-task and $u_b$ is the required CPU cycles to compute $1$-bit of offloaded sub-task at each server $b$. Because each node in the network is mobile, the total communication time $T^{\mathrm{tran}}_{d_{\mu}, b}$ and computing time $T^{\mathrm{comp}}_{d_{\mu}, b}$ cannot exceed the maximum CTE and respective satellite connection time\footnote{Each satellite has different moving speed, thus provide different coverage time. However, CNS provides services in each time slot $n$.} between $\mu$ and $b$. We can denote it as $T^{\mathrm{max}}_{d_{\mu}, b}$ and define a constraint for it as:
\begin{equation}
    T^{\mathrm{tran}}_{d_{\mu}, b} + T^{\mathrm{comp}}_{d_{\mu}, b} \leq  T^{\mathrm{max}}_{d_{\mu}, b}.
\end{equation}
The aggregate of allotted computing resources cannot thus exceed the maximum computing capacity, which is defined as:
\begin{equation}
    \sum_{\forall d \in \mathcal{D}} \sum_{\forall \mu \in \mathcal{E}_l}  x^{b}_{d_{\mu}}\beta_{d_{\mu}}^b \leq \varrho_b.
\end{equation}
Now we can define the computing power rental price for each satellite. Let's consider that the rental computing unit price for LMS, CubeSat, and CNS is defined as $\chi^{\mathrm{comp}}_l$, $\chi^{\mathrm{comp}}_c$, and $\chi^{\mathrm{comp}}_h$, respectively. Thus, the computing price of task $d_{\mu}$ processing by the respective source can be defined as follows:
\begin{equation}
\begin{cases}
    P^{\mathrm{comp}}_{d_{\mu}, c} = \chi^{\mathrm{comp}}_c  \beta_{d_{\mu}}^b  \omega_b,~\forall b \in \mathcal{C} & \mathrm{CubeSat~ price},\\
    P^{\mathrm{comp}}_{d_{\mu}, b} = \chi^{\mathrm{comp}}_l  \beta_{d_{\mu}}^b  \omega_b,~\forall b \in \mathcal{L} & \mathrm{LMS~ price},\\
    P^{\mathrm{comp}}_{d_{\mu}, h} = \chi^{\mathrm{comp}}_h  \beta_{d_{\mu}}^b  \omega_b,~\forall b \in \mathcal{H}, & \mathrm{CNS~price}.
\end{cases}
\end{equation}
where the unit of processing computing task is considered as cyc/sec/\$. We believe that the unit price of CNS is greater, followed by the price of LMC, and finally the price of CubeSats, because of their proximity and communication distance, which can be defined as $\chi^{\mathrm{comp}}_h >\chi^{\mathrm{comp}}_l > \chi^{\mathrm{comp}}_c$. 

\section{Problem Formulation}
\label{prob_form}
In this section, we formulate our problem using system modeling from the previous section. By utilizing the sub-task communication and computing model, we can define the service time and price of each sub-task $d_{\mu}$, which are indicated as $T^{\mathrm{ser}}_{d_{\mu}}$ and $P^{\mathrm{ser}}_{d_{\mu}}$, respectively as below:
\begin{equation}
    T^{\mathrm{ser}}_{d_{\mu}} = \sum_{\forall b \in \mathcal{B}_{\mu}} x^b_{d_{\mu}} (T^{\mathrm{tran}}_{d_{\mu}, b} + T^{\mathrm{comp}}_{d_{\mu}, b}),
\end{equation}
\begin{equation}
    P^{\mathrm{ser}}_{d_{\mu}} = \sum\limits_{\forall b \in \mathcal{B}_{\mu}} x^b_{d_{\mu}} (P^{\mathrm{tran}}_{d_{\mu}, b} + P^{\mathrm{comp}}_{d_{\mu}, b}).
\end{equation}
We consider that each task $d$'s service time can be calculated as all the sub-tasks $d_{\mu} \in \mathcal{D}$ mean service time as:
\begin{equation}
    T^{\mathrm{mean}}_{d} = \sum_{\forall d_{\mu} \in d} \frac{T^{\mathrm{ser}}_{d_{\mu}}}{\|d\|},~\forall d \in \mathcal{D}.
\end{equation}
Likewise, we consider that each task $d$'s service price can be calculated as all the sub-tasks $d_{\mu} \in \mathcal{D}$ mean service price as:
\begin{equation}
     P^{\mathrm{mean}}_{d} = \sum_{\forall d_{\mu} \in d} \frac{P^{\mathrm{ser}}_{d_{\mu}}}{\|d\|},~\forall d \in \mathcal{D}. 
\end{equation}
Based on the above derivation, we can define our two objectives in the next definitions:
\begin{definition}[Mean Service Time (MST)]
MST is explained as the aggregation of tasks' service time divided by the entire number of tasks, and that can rationally measure network efficiency.
\begin{equation}
    \eta_1 = \sum_{\forall d \in \mathcal{D}} \frac{T^{\mathrm{mean}}_{d}}{\|D\|} = \sum_{\forall d \in \mathcal{D}} \sum_{\forall d_{\mu} \in d} \frac{T^{\mathrm{ser}}_{d_{\mu}}}{\|d\|  \|D\|}.
\end{equation}
\end{definition}
\begin{definition}[Mean Service Price (MSP)]
MSP is explained as the aggregation of tasks' service prices divided by the entire number of tasks, and that can rationally measure network costs.
\begin{equation}
    \eta_2 = \sum\limits_{\forall d \in \mathcal{D}} \frac{P^{\mathrm{mean}}_{d}}{\|D\|} = \sum\limits_{\forall d \in \mathcal{D}} \sum\limits_{\forall d_{\mu} \in d} \frac{P^{\mathrm{ser}}_{d_{\mu}}}{\|d\|  \|D\|}.
\end{equation}
\end{definition}
To minimize both objectives at the same time, we can define the objective function as the weighted sum of MST and MSP, which can be calculated as below:
\begin{equation}
    \eta = \alpha_1 \eta_1 + \alpha_2  \eta_2 = \sum_{\forall d \in \mathcal{D}} \sum_{\forall d_{\mu} \in d} \frac{\alpha_1  T^{\mathrm{ser}}_{d_{\mu}} + \alpha_2 P^{\mathrm{ser}}_{d_{\mu}} }{\|d\| \|D\|},
\end{equation}
where weights for MST and MSP are given as $\alpha_1$ and $\alpha_2$, respectively, and their summation should not increase by $1$, i.e., $\alpha_1 + \alpha_2 = 1$. After that, we can define our proposed optimization problem for data-driven ITS task offloading. Mathematically, it can be defined as below:
\begin{mini!}|s|[2]<b>
		{\substack{\textbf{X}, \textbf{Y}, \boldsymbol{\beta},  \boldsymbol{\omega} }}
		{\eta = \sum_{\forall \mu \in \mathcal{D}} \frac{\alpha_1 T^{\mathrm{ser}}_{d} + \alpha_2  P^{\mathrm{ser}}_{d}}{\|D\|}} {\label{opt:P1}}{\textbf{P1:}}
		\addConstraint{\sum_{\forall b \in \mathcal{B}} x^{b}_{d_{\mu}} = 1, ~\forall d \in \mathcal{D},~ \forall \mu \in \mathcal{E} {\label{C1}}}
		\addConstraint{\sum_{\forall d \in \mathcal{D}} \sum_{\forall \mu \in \mathcal{E}_l} x_{d_{\mu}}^b y^b_{d_{\mu}} \leq 1, ~\forall b \in \mathcal{B} {\label{C2}}}
		\addConstraint{\sum_{\forall d \in \mathcal{D}} \sum_{\forall \mu \in \mathcal{E}_l}  x^{b}_{d_{\mu}}\beta^b_{d_{\mu}} \leq \varrho_b, ~\forall b \in \mathcal{B} {\label{C3}}}
		\addConstraint{  x^{b}_{d_{\mu}} \{T^{\mathrm{tran}}_{d_{\mu}, b} + T^{\mathrm{comp}}_{d_{\mu}, b}\} \leq T^{\mathrm{max}}_{d_{\mu}, b},~ \forall b \in \mathcal{B}  {\label{C4}}}
		\addConstraint{  \sum_{\forall d \in \mathcal{D}} \sum_{\forall \mu \in \mathcal{E}_c}  x^{b}_{d_{\mu}} \leq 1, ~\forall b \in \mathcal{C}_l {\label{C5}}}
		\addConstraint{ x^{b}_{d_\mu} \in \{0,1\}, ~\forall b \in \mathcal{B}_{\mu}, ~\forall d_{\mu} \in d,~\forall d \in \mathcal{D} {\label{C6}}}
		\addConstraint{ \omega^h_{d_{\mu}} \in R^+, ~\forall d_{\mu} \in d,~\forall d \in \mathcal{D}. {\label{C7}}}
        \addConstraint{ y^b_{d_{\mu}} \leq Y^{\mathrm{th}}, ~\forall b \in \mathcal{B} {\label{C8}}}
        \addConstraint{ \beta^b_{d_{\mu}} \leq \beta^{\mathrm{th}}, ~\forall b \in \mathcal{B}. {\label{C9}}}
\end{mini!}
Note that a set $\mathcal{X}=\{x^b_{d_{\mu}}\}$ denotes the binary ($0$-$1$) integer variables for offloading decisions, $\mathcal{Y}=\{y_{d_{\mu}}\}$ denotes communication bandwidth resources allocation variables, $\mathcal{\boldsymbol{\beta}}=\{\beta_{d_{\mu}}\}$ denotes computing resources allocation variables for LMSs and CubeSats, and 
$\mathcal{\boldsymbol{\omega}}=\{\omega^h_{d_{\mu}}\}$ denotes CNS computing power allocation, which should be positive continuous variables. The constraint (\ref{C1}) specifies that the summation of the associated CTE shall be one at each time slot. Constraint (\ref{C2}) reflects the requirement that the total wireless bandwidth allocation ratio is less than or equal to one. The constraint (\ref{C3}) guarantees that the allocated LMS and CubeSats computing resources do not exceed the threshold. The constraint (\ref{C4}) guarantees that the communication time between the CTE and the related satellite is shorter than the maximum permitted time. Constraint (\ref{C5}) indicates that only those CubeSats that are already in the neighborhood of LMS $l$ can give service to CTE. Constraint (\ref{C6}) indicates that CTEs can only associate with one satellite at a time. Constraint (\ref{C7}) ensures that each CNS $h$ has enough computing resources. Constraint (\ref{C8}) and (\ref{C9}) ensure that communication bandwidth and computing resources must remain within the budget. We consider that a weighted objective function should have a positive value to achieve balance. The defined optimization problem is an MINLP and NP-hard problem due to non-linearity in an objective function and the availability of binary variables that are difficult to handle directly. To address this proposed problem, we will divide it into two stages in the next section.

\section{Proposed Solution Algorithm}
\label{prop_algo}
We develop the data-driven ITS tasks offloading solution in two components: sub-task offloading and resource allocation. For sub-task offloading, we first create a Co-MAPPO DRL with an attention algorithm on each satellite. After that, the theoretically optimal solutions for resource allocation are obtained using decomposition and convex optimization.
\subsection{First Stage: Cooperative Multi-Agent Proximal Policy Optimization DRL }
In this part, we will deal with how to optimize the satellite offloading decision variables $\boldsymbol{X}$ at each time slot $n$. Thus, given the initial\footnote{We consider that each decision variable is defined in-bounds and provides a definite value of the objective function.} communication bandwidth resource allocation $\boldsymbol{Y}$, computing resource allocation $\boldsymbol{\beta}$, and CNS computing power allocation ${\omega}^h_{d_{\mu}}$, $\textbf{P1}$ is converted into $\textbf{P1.1}$. Mathematically, it can be described as follows:
\begin{mini!}|s|[2]<b>
		{\substack{\textbf{X} }}
		{\eta = \sum_{\forall \mu \in \mathcal{D}} \frac{\alpha_1 T^{\mathrm{ser}}_{d} + \alpha_2  P^{\mathrm{ser}}_{d}}{\|D\|}} {\label{opt:P1.1}}{\textbf{P1:1}}
		\addConstraint{\sum_{\forall b \in \mathcal{B}} x^{b}_{d_{\mu}} = 1, ~\forall d \in \mathcal{D},~ \forall \mu \in \mathcal{E} {\label{C1.1}}}
		\addConstraint{  x^{b}_{d_{\mu}} \{T^{\mathrm{tran}}_{d_{\mu}, b} + T^{\mathrm{comp}}_{d_{\mu}, b}\} \leq T^{\mathrm{max}}_{d_{\mu}, b},~ \forall b \in \mathcal{B}  {\label{C1.2}}}
		\addConstraint{  \sum_{\forall d \in \mathcal{D}} \sum_{\forall \mu \in \mathcal{E}_c}  x^{b}_{d_{\mu}} \leq 1, ~\forall b \in \mathcal{C}_l {\label{C1.3}}}
		\addConstraint{ x^{b}_{d_\mu} \in \{0,1\}, ~\forall b \in \mathcal{B}_{\mu}, ~\forall d_{\mu} \in d,~\forall d \in \mathcal{D} {\label{C1.4}}}	
\end{mini!}
Due to the non-convexity caused by constraints (\ref{C1.1}), (\ref{C1.3}) and (\ref{C1.4}), it can be shown that problem (\ref{opt:P1.1}) is non-convex and non-linear. Furthermore, because the objective function in (\ref{opt:P1.1}) has long-term accumulation characteristics, we can convert this optimization problem to a sequential decision-making problem and use the MADRL method to solve the specific problem. Co-MAPPO DRL in ITS networks refers to the learning problem in which multiple satellites can learn decision policy at the same time. Each satellite can only gather local observations from the environment. Because the actions of one satellite might affect the performance of other satellites, thus this study proposes a global incentive to lead the decentralized DRL. To address the problem (\ref{opt:P1.1}), a partially observable Markov decision process (POMDP) model with state space, action space, and reward is initially developed as below:

\textbf{Network State Space:} Due to the restricted CTE-CubeSat range, it is believed that each CTE has access to no more than $m$ CubeSats. The network status of the current pending sub-task $d_{\mu}$ at time slot $n$ is thus explained as a multidimensional array, which can be defined as:
    \begin{equation}
        \boldsymbol{s}_{d_{\mu}}(n) = [M_{d_{\mu}}, \nu_{d_{\mu}}, M^{\mathrm{tot}}, M^{\mathrm{load}}, \zeta_c, \eta_c, \zeta_{l},\eta_{l}],~\forall c,l\in \mathcal{B}
    \end{equation}
where $M_{d_{\mu}}$ is the required task data memory, $\nu_{d_{\mu}}$ is the required task computing power, $M^{\mathrm{tot}}$ is the full data memory for queuing sub-tasks that are awaiting association, and $M^{\mathrm{load}}$ is the tasks that have already been offloaded to respective server $b$. Moreover, the observation space of the $b$-th satellite in time slot $n$ can be characterized as:

\begin{equation}
        z_{b}(n) = O_{b}(\boldsymbol{s}_{d_{\mu}}(n)), ~b \in \mathcal{L} \cup \mathcal{C} \cup \mathcal{H},
    \end{equation}
where observation function $O(\boldsymbol{s}_{d_{\mu}})$ is used by the agent $b \in \{1, \cdots ,B\}~ \forall b \in \mathcal{L} \cup \mathcal{C} \cup \mathcal{H}, $ to gain its partial observation.

\textbf{Network Action Space:} We can define actions as the collection of accessible candidate computing satellites for awaiting task $d_{\mu}$. As demonstrated in equation (\ref{action_array}), one-hot encoding is used to express the action $\boldsymbol{a}_{d_{\mu}}(n)$ in each time slot $n$, which is a binary array represented in $b$-dimensional as:
    \begin{equation}
        \boldsymbol{a}_{d_{\mu}}(n) = [a^c_{d_{\mu}}, a^l_{d_{\mu}}, a^{h}_{d_{\mu}}],~ \forall c,l,h\in \mathcal{B}. \label{action_array}
    \end{equation}

\textbf{Network Reward Function:} The reward function's core idea is that the shorter the action's service time and price, the greater the reward. If the sub-task is performed successfully during the satellite access time, the reward is characterized as the product of the two factors: the inverse of the weighted sum of service time and sub-task price; and a constant $\Gamma_1$. Otherwise, the reward is a negative value of $\Gamma_2$, indicating a penalty. Mathematically, it can be represented as below:
    \begin{equation}
            r_{d_{\mu}}(n)=
        \begin{cases}
            \frac{\Gamma_1}{\alpha_1 T^{\mathrm{ser}}_{d_{\mu}} + \alpha_2  P^{\mathrm{ser}}_{d_{\mu}}}, & T^{\mathrm{mean}}_{d_{\mu},b} - T^{\mathrm{max}}_{\mu, b} \leq 0,\\
            -\Gamma_2, & \mathrm{otherwise}
        \end{cases}
    \end{equation}
where $\Gamma_1$ and $\Gamma_2$ are two predetermined constants that are positive, and $b$ refers to the chosen satellite. It should be observed that the reward becomes available only once all pending sub-tasks have been accomplished. The cumulative discounted rewards can be calculated using a discount rate $\gamma \in [0, 1)$ as follows:
    \begin{equation}
        R(n) = \sum_{k=0}^{\infty} \gamma^k r_{d_{\mu}}(n+k+1). 
    \end{equation}
Let $\boldsymbol{\pi}=\{\pi_b|b\in\mathcal{B}\}$ be the multiple agents' combined policy, and $V^{\pi}$ signify the state-value function, then:
\begin{equation}
    V^{\boldsymbol{\pi}}(s(n)) =  \mathbb{E}_{a(n), s(n+1), \cdots }[R(n)|s(n)].
\end{equation}
The answer to offloading decision problem can be applied to determining the best way to maximize the expected total of discounted rewards for the starting state $s(0)$. As a result, a Co-MAPPO DRL optimization problem is phrased as follows:
\begin{equation}
\begin{aligned}
    \boldsymbol{\pi}^{*}    & =  \arg  \underset{\boldsymbol{\pi}}{\mathrm{max}} ~  \mathbb{E}_{s(0)\sim\rho_s(s(0)) } [V^{\boldsymbol{\pi}} (s(0))], \\
                            & =  \arg \underset{\boldsymbol{\pi}}{\mathrm{max}} ~ \kappa (\boldsymbol{\pi}).  \label{opt_DRL}
\end{aligned}
\end{equation}
\subsubsection{Attention Mechanism for Co-MAPPO DRL}
We examine an attention mechanism for CO-MAPPO DRL with the idea that agents (satellites) would be able to focus on information relevant to their reward while learning the strategies of competing agents rather than receiving the same content. Attention would be essentially a query mapping to a series of (key-value) pairs that enables each agent to evaluate the relevance of competing agents to itself by querying competing agents' occurring states and produced actions and incorporating this knowledge into their action value approximation function i.e., $Q$-value function \cite{pmlr-v97-iqbal19a, sat_attention}. The critic gets observations $\boldsymbol{z} = (z_1, \cdots , z_B)$ and actions $\boldsymbol{a} = (a_1, \cdots ,a_B)$, for all agents indexed by $b \in \{1, \cdots , B\}$ in order to evaluate the $Q$-value function $Q^{\theta}_b(z, a)$ for the agent $b$. All agents except $b$ are represented as $^{-}b$, and this set is indexed with $i$. The information from another agent $i$'s is taken into account via the $Q$-value function $Q^{\theta}_b(z, a)$, which may be represented as:
\begin{equation}
   Q^{\theta}_b (z, a) = f_b (g_b (z_b, a_b), \psi_b), \label{attention_estimation}
\end{equation}
where $f_b$ denotes the two-layer multi-layer perceptron (MLP), $g_b$ is one-layer MLP which represents an embedding function and $\psi_b$ is a weighted sum of encoding values of the other agents, representing their contributions. The following is the procedure for calculating weights $\phi_i$, $i = \{1, 2, \cdots, b-1, b + 1, \cdots, B\}$: initially, all other agents' actions and states are encoded using the embedding function as $e_i = g_i(z_i, a_i)$. The agent $b$'s state and action are then represented as $e_b = p_b(z_b, a_b)$. The similarity between $e_i$ and $e_b$ is then compared using a bilinear mapping (i.e., the query-key system), and the resulting values are normalized through the softmax layer to obtain the weight $\phi_i$. A simple transformation may be used to acquire the encoding value $\nu_i$. The states and actions of competing agents are first encoded using an embedded function, $e_i$, and then a linear transformation is performed using the shared matrix $V$:
\begin{equation}
    \psi_b = \sum_{i \neq b}^{B} \phi_i \nu_i = \sum_{i \neq b}^{B} \phi_i \Psi(Vg_i(z_i, a_i)), \label{attention_code}
\end{equation}
where $\phi_i \propto \exp{(e_i^{\mathrm{T}}W_k^{\mathrm{T}} W_{q}  e_b)}$, $\Psi(\cdot)$ is a nonlinear transformation function, $W_k$ transforms $e_i$ to "key", and $W_q$ transforms $e_b$ into a "query". To avoid vanishing gradients, the matching is scaled by the dimensions of these two matrices.
\subsubsection{Learning Procedure of Co-MAPPO DRL with attention}
\begin{figure*}[t]
    \centering
    \includegraphics[width=0.7\textwidth, height=4in]{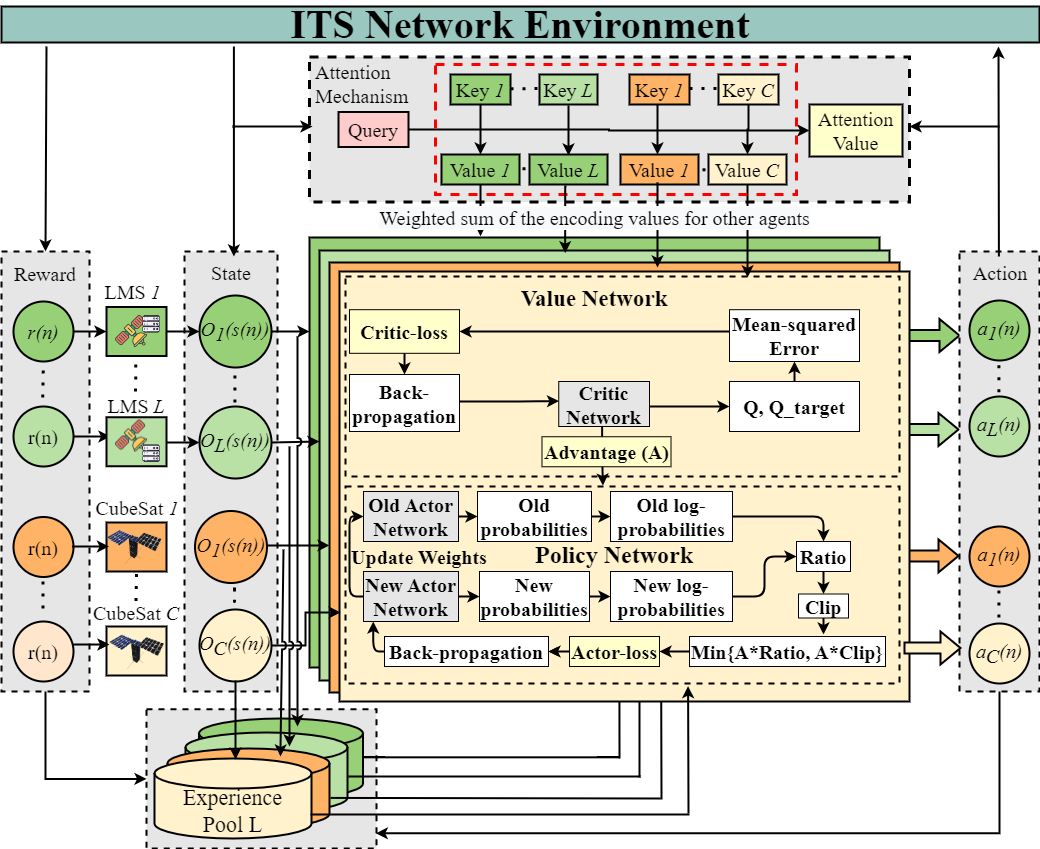}
    \caption{Proposed framework of Co-MAPPO DRL with attention mechanism for ITS data-driven task offloading management.}
    \label{solution_flowchart}
    \vspace{-0.1in}
\end{figure*}

Because the environment of an ITS network is rapidly changing, we use the PPO strategy to deal with the MADRL problem (\ref{opt_DRL}) to provide a stable policy learning procedure. In comparison to DQN, PPO, being one of the policy gradient techniques, is less sensitive to hyper-parameters in the training phase. Each satellite $b$ in the proposed system functions as an agent with a policy network $\boldsymbol{\theta}_b$ and a value network $\boldsymbol{\tau}_b$. Maximizing $\kappa(\boldsymbol{\pi})$ with given $\boldsymbol{\pi}^{\mathrm{old}}$ is identical to maximizing $\mathbb{E}_{\boldsymbol{\pi}}[A^{\boldsymbol{\pi}^{\mathrm{old}}}(s(n),a(n))]$ \cite{schulman2017proximal}, where $A^{\boldsymbol{\pi}}(s(n), a(n))=Q^{\boldsymbol{\pi}}(s(n), a(n)) - V^{\boldsymbol{\pi}}(s(n))$ is the advantage function and the state-action value function is as follows:
\begin{equation}
    Q^{\boldsymbol{\pi}}(s(n), a(n))=\mathbb{E}_{a(n+1),s(n+1), \cdots}[R(n)|(s(n), a(n))]
\end{equation}
Furthermore, a clip function may be used to approximate $\mathbb{E}_{\boldsymbol{\pi}}[A^{\boldsymbol{\pi}^{\mathrm{old}}}(s(n),a(n))]$, allowing optimization issue (\ref{opt_DRL}) to be changed into:
\begin{multline}
    \underset{\theta=\{\theta_b|b\in \mathcal{B} \}}{\mathrm{max}} \mathbb{E}_{\boldsymbol{\pi}^{\mathrm{old}}} \Big[ \sum_{b \in \mathcal{B}} \mathrm{min}\{\Phi(\theta_b)A^{\boldsymbol{\pi}^{\mathrm{old}}}(s(n),a(n)),\\
    \mathrm{clip}(\Phi(\theta_b), \epsilon )A^{\boldsymbol{\pi}^{\mathrm{old}}}(s(n),a(n)) \}\Big], \label{PPO}
\end{multline}
where $\Phi(\boldsymbol{\theta_b}) = \frac{\boldsymbol{\pi}_b(a_b|z_b;\boldsymbol{\theta})}{\boldsymbol{\pi}^{\mathrm{old}}_b(a_b|z_b;\boldsymbol{\theta}^{\mathrm{old}})}$ is $b$-th satellite probability ratio. The clip function is used to keep the value of $\Phi(\boldsymbol{\theta}_b)$ inside the range $[1-\epsilon,1+\epsilon]$ limiting the policy update to a short range. When the actor network's parameters are updated, the sampled action is sent to the critic network, which then calculates the action's estimated value using the attention technique. The actor can be modified using a strategy gradient technique based on the estimated value \cite{pmlr-v97-iqbal19a}. With saved samples, the expectation of the advantage function $A^{\boldsymbol{\pi}^{\mathrm{old}}}$ in (\ref{PPO}) can be approximated during the training phase. In other words, the following gradient is used to update the policy:
\begin{multline}
    \Delta \boldsymbol{\theta}_b = \nabla_{\boldsymbol{\theta}_b} \hat{\mathbb{E}} \Big[ \mathrm{min}\{\Phi(\boldsymbol{\theta}_b)A_b(s(n),a(n)),\\
    \mathrm{clip}(\Phi(\boldsymbol{\theta}_b), \epsilon )A_b(s(n),a(n)) \}\Big],  \label{actor_gradient}
\end{multline}
where $A_b(s(n), a(n))$ is generalized advantage estimation (GAE) and can be defined as:
\begin{equation}
    A_b(s(n),a(n)) = \hat{Q}(s(n),a(n)) - V_{\boldsymbol{\tau}^{\boldsymbol{\mathrm{old}}}_b}(s(n)), \label{advantage}
\end{equation}
where $\hat{Q}_b(s(n), a(n))$ can be estimated by using the k-step bootstrapping linear combination as:
\begin{equation}
    \hat{Q}(s(n),a(n))) = \sum_{k=n}^{\infty} (\Xi \upsilon)^{k-n} \delta(n) + Q_{\boldsymbol{\tau}^{\mathrm{old}}_b}(s(n), a(n)), \label{Q_value}
\end{equation}
where $\delta(n)$ is the temporal difference (TD) \cite{sutton2018reinforcement} and could be defined as:
\begin{equation}
    \delta(n) = \{r(n) + \Xi Q_{\boldsymbol{\tau}_b} (s(n+1), a(n+1)) - Q_{\boldsymbol{\tau}_b^{\mathrm{old}}}(s(n),a(n))\}.
\end{equation}
Because each satellite $b$ seeks to estimate the same combined state-action value function, the value network's loss function can be calculated as follows:
\begin{equation}
    L^{\mathrm{value}}(n, \boldsymbol{\tau}_b) = (\hat{Q}_b(s(n),a(n)) - Q_{\boldsymbol{\tau}^{\mathrm{old}}_b}(s(n),a(n)))^2.
\end{equation}
As a result, an optimization problem is devised to train the value network as:
\begin{multline}
    \underset{\boldsymbol{\tau}=\{\boldsymbol{\tau}_b|b\in \mathcal{B} \}}{\mathrm{min}} \mathbb{E} \Big[ \sum_{b \in \mathcal{B}} (\hat{Q}_b(s(n),a(n)) - Q_{\boldsymbol{\tau}^{\mathrm{old}}_b}(s(n),a(n)))^2 \Big].\label{value_update}
\end{multline}
The gradient descent approach can be utilized to address this problem (\ref{value_update}). The gradient value for each satellite could be:
\begin{multline}
    \Delta \boldsymbol{\tau}_b = \nabla_{\boldsymbol{\tau}_b} \hat{\mathbb{E}} \Big[ (\hat{Q}_b(s(n),a(n)) - Q_{\boldsymbol{\tau}^{\mathrm{old}}_b}(s(n),a(n)))^2\Big],  \label{critic_gradient}
\end{multline}
where parameters $\boldsymbol{\theta}$ and $\boldsymbol{\tau}$ can be changed until the loss functions of the policy network and the value network converge, according to (\ref{actor_gradient}) and (\ref{critic_gradient}), respectively. The centralized training approach is used to update the critic network. Each agent's local state and actions are initially dispatched to the attention framework, and then the proportional sum of coded values is produced using (\ref{attention_code}), guided by estimation utilizing (\ref{attention_estimation}), and lastly, the TD technique is utilized to fine-tune the critic's parameters with the lowest loss function. To tackle the credit assignment problem, we employ the counterfactual baseline provided by \cite{foerster2018counterfactual}. We estimate the action-value function $Q^{\boldsymbol{\pi}_{\mathrm{old}}}$ using a centralized critic $Q_{\tau_b}(s(n),\boldsymbol{a}(n))$. We use the superscript $-b$ to represent joint quantities over satellites other than a given satellite $b$, for example, a joint action other than the satellite $b$ is an $\boldsymbol{a}^{-b}(n)$. The advantage function is then derived for each satellite $b$ by comparing the $Q$-value assessed by the critic for the executed action $a^b(n)$ to a counterfactual baseline that marginalizes out $a^{b}(n)$ while keeping the actions of other satellites constant as:
\begin{equation}
    A^{b}(s(n),\boldsymbol{a}(n))  = \hat{Q}^{b}(s(n), \boldsymbol{a}(n)) - b(s(n), \boldsymbol{a}^{-b}(n)), \label{central_adv}
\end{equation}
where 
\begin{equation}
    b(s(n), \boldsymbol{a}^{-b}(n)) = \sum_{a^b} \pi^{b}_{\mathrm{old}} (a^b|z^b(n)) Q_{\tau^b}(s(n), (\boldsymbol{a}^{-b}(n), a^b)) 
\end{equation}
refers to the counterfactual baseline, and $\hat{Q}^b(s(n), \boldsymbol{a}(n))$ denotes the estimation of $Q^{\boldsymbol{\pi}_{\mathrm{old}}}(s(n), \boldsymbol{a}(n))$. Moreover, $\hat{Q}^b(s(n), \boldsymbol{a}(n))$ can be characterized by truncated TD as:
\begin{multline}
  \hat{Q}^b(s(n), \boldsymbol{a}(n)) = Q_{\Bar{\tau}^b}(s(n), \boldsymbol{a}(n)) + \delta(n) + \\(\Xi \nu)\delta(n+1) + \cdots + (\Xi \nu)^T \delta(n), 
\end{multline}
where error of TD is $\delta(n) = r(n) + \Xi Q_{\Bar{\tau}^b}(s(n+1), \boldsymbol{a}(n+1)) - Q_{\Bar{\tau}^b}(s(n), \boldsymbol{a}(n))$. Thus, (\ref{central_adv}) computes a distinct advantage function for each satellite that uses the centralized critic to infer counterfactuality in which only the satellite $b$'s behavior changes. Despite the fact that each $\hat{Q}^b(s(n),\boldsymbol{a}(n))$ is computed using a distinct critic, they all estimate the same joint action-value function $Q^{\boldsymbol{\pi}_{\mathrm{old}}}(s(n), \boldsymbol{a}(n))$. Fig. \ref{solution_flowchart} depicts the proposed Co-MAPPO DRL framework.
\subsubsection{Description of Learning Procedure of Co-MAPPO DRL with attention}
The Co-MAPPO algorithm is a distributed learning approach that utilizes DRL to optimize decision-making processes in the SATs for ITS environment. Fig. \ref{alg:MAPPO} presents a conceptual diagram that provides an in-depth analysis of the Co-MAPPO DRL learning process with an attention mechanism. In this approach, each satellite, including the LMS and CubeSat, is treated as a learning agent that receives rewards and local state observations from the network controller. The agents learn policies with the aid of the MAPPO algorithm, which employs two neural networks, namely the policy and value networks, for decision-making and evaluation, respectively. The transitional data, such as states, actions, rewards, and next states, are stored in an experienced pool for neural network updates. To reduce computational complexity, the attention mechanism assigns weightage only to inputs that provide relevant information to the agent. Thus, the proposed approach provides a promising solution for optimizing decision-making in the ITS network environment.
\subsection{Second Stage: Decomposition \& Convex Optimization for Subproblems}
The value of $\boldsymbol{X}$ is calculated in advance using the Co-MAPPO DRL. Thus optimization problem (\ref{opt:P1}) is concerned with resource allocation, which is reformulated as:
\begin{mini!}|s|[2]<b>
		{\substack{\textbf{Y}, \boldsymbol{\beta},  \boldsymbol{\omega} }}
		{\eta = \sum_{\forall d \in \mathcal{D}} \frac{\alpha_1 T^{\mathrm{ser}}_{d} + \alpha_2  P^{\mathrm{ser}}_{d}}{\|D\|}} {\label{opt:P1.2}}{\textbf{P1.2:}}
		\addConstraint{\sum_{\forall d \in \mathcal{D}} \sum_{\forall \mu \in \mathcal{E}_l} x_{d_{\mu}}^b y_{d_{\mu}} \leq 1, ~\forall b \in \mathcal{B} {\label{C1.2.1}}}
		\addConstraint{\sum_{\forall d \in \mathcal{D}} \sum_{\forall \mu \in \mathcal{E}_l}  x^{b}_{d_{\mu}}\beta_{d_{\mu}} \leq \varrho_b, ~\forall b \in \mathcal{B} {\label{C2.1.2}}}
		\addConstraint{ \omega^h_{d_{\mu}} \in R^+, ~\forall d_{\mu} \in d,~\forall d \in \mathcal{D}. {\label{C1.2.3}}}
        \addConstraint{ y^b_{d_{\mu}} \leq Y^{\mathrm{th}}, ~\forall b \in \mathcal{B} {\label{C1.2.4}}}
        \addConstraint{ \beta^b_{d_{\mu}} \leq \beta^{\mathrm{th}}, ~\forall b \in \mathcal{B}. {\label{C1.2.5}}}
\end{mini!}
The variables $\boldsymbol{Y}$, $\boldsymbol{\beta}$ and $\boldsymbol{\omega}$ in problem (\ref{opt:P1.2}) are shown to be independent of one another. Because the variables do not overlap, the three constraints (\ref{C1.2.1})-(\ref{C1.2.3}) are separable. Therefore, we can decompose three subproblems:
\subsubsection{Communication Bandwidth Resource Allocation} The first subproblem concerning the variables $\boldsymbol{Y}$ is the communication bandwidth allocation for CTE task offloading, which is expressed as:
\begin{mini!}|s|[2]<b>
		{\substack{\boldsymbol{Y}}}
		{\Upsilon_2 =  \sum_{\forall d \in \mathcal{D}} \sum_{\forall d_{\mu} \in d} \frac{\alpha_1 x^{b}_{d_{\mu}} T^{\mathrm{tran}}_{d_{\mu}, b} } {\|d\|\|D\|} } {\label{opt:P1.2.b}}{\textbf{P1.2.a:}}
		\addConstraint{\sum_{\forall d \in \mathcal{D}} \sum_{\forall \mu \in \mathcal{E}_l} x_{d_{\mu}}^b y_{d_{\mu}} \leq 1, ~\forall b \in \mathcal{B} {\label{C1.2.b1}}}
\end{mini!}
Since the variables connected to each CubeSats, LMSs, and CNS are independent of one another, the subproblem (\ref{opt:P1.2.b}) may be decomposed into numerous simple problems, each of which is only related to one server $b$, as follows:
\begin{mini!}|s|[2]<b>
		{\substack{\boldsymbol{Y}_b}}
		{\Upsilon^b_2 =  \sum_{\forall d_{\mu} \in \mathcal{D}^b} \frac{\alpha_1 x^{b}_{d_{\mu}} T^{\mathrm{tran}}_{d_{\mu}, b} } {\|d\|\|D\|} } {\label{opt:P1.2.b.1}}{\textbf{}}
		\addConstraint{\sum_{\forall d_{\mu} \in \mathcal{D}^b}  x_{d_{\mu}}^b y_{d_{\mu}} \leq 1,  {\label{C1.2.b.1}}}
\end{mini!}
where $D^b$ refers to all sub-tasks within server $b$'s coverage and $\boldsymbol{Y}_b$ denotes the variables in $\boldsymbol{Y}$ connected with server $b$. The problem (\ref{opt:P1.2.b.1})'s objective function is convex, while the constraint (\ref{C1.2.b.1}) is linear. Therefore, we have a convex problem (\ref{opt:P1.2.b.1}) and can find the optimal solution with Karush–Kuhn–Tucker (KKT) conditions, i.e., stationarity condition, complementary slackness, and dual feasibility, respectively \cite{boyd2004convex} as:
\begin{equation}
    \begin{aligned}
        \nabla_{\boldsymbol{Y}_l} \Upsilon^b_2 + \iota_b \nabla_{\boldsymbol{Y}_b}\Big( \sum_{\forall d_{\mu} \in \mathcal{D}^b}  x_{d_{\mu}}^b y_{d_{\mu}} - 1 \Big) & = 0, \\
        \iota_b\Big( \sum_{\forall d_{\mu} \in \mathcal{D}^b}  x_{d_{\mu}}^b y_{d_{\mu}} - 1 \Big) & = 0, \quad \iota_b & \geq 0.
    \end{aligned}
\end{equation}
The optimal closed-form solution of wireless bandwidth allocation for sub-task $d_{\mu}$ can be determined by solving the set of equations as below:
\begin{equation}
    \begin{aligned}
        y^*_{d_{\mu}} & = \frac{x^{b}_{d_{\mu}}\sqrt{\vartheta_{d_{\mu}}}}{\sum_{\forall d_{\mu} \in \mathcal{D}^b} x^{b}_{d_{\mu}}\sqrt{\vartheta_{d_{\mu}}}}, \\ \label{KKT_sol_comm}
         & \mathrm{where~~} \vartheta_{d_{\mu}} = \frac{\alpha_1 M_{d_{\mu}}}{\zeta_b  \|d\| \log_2(1+\frac{p_{\mu} g_{\mu, b} }{N_0})},~\forall d_{\mu} \in \mathcal{D}^b. 
    \end{aligned}
\end{equation}
\begin{algorithm}[t!]
\small
\caption{\strut Co-MAPPO DRL and Convex Optimization based ITS Data Management (\ref{opt:P1.2})} 
\label{alg:MAPPO}
\begin{algorithmic}[1]
\STATE{Initialize critic $Q_{\boldsymbol{\tau}^b}$ and actor $\pi^b$ with $\boldsymbol{\theta}^b$, $\forall b \in \mathcal{L\cup C}$.}
\STATE{ Initialize the current policies $\pi^b_{\mathrm{old}}$ with $\boldsymbol{\theta}^b_{\mathrm{old}} \leftarrow \boldsymbol{\theta}^b$, and the critic $Q_{\bar{\boldsymbol{\tau}}^b}$ with $\bar{\boldsymbol{\tau}}^b \leftarrow  \boldsymbol{\tau}^b$. }
\STATE{ Initialize a experience pool $\boldsymbol{L}$}
\FOR{episode$=1,2,...,E$}
\STATE {Give initial bandwidth resource $\boldsymbol{Y}$, computing resource $\boldsymbol{\beta}$, and CNS computing power ${\omega}^h_{d_{\mu}}$.}
\STATE{Solve $\textbf{P1}$ which is converted into $\textbf{P1.1}$ with CO-MAPPO DRL.}
\FOR{time slot~$n=1,2,...,N$}
\STATE{ $\boldsymbol{s}_{d_{\mu}}$ = initiate state}
\STATE{ Each satellite agent obtain its observations $O_b(\boldsymbol{s(n)})$}
\STATE{ Each satellite agent obtains encoding state values for other agents from attention mechanism $\psi_b$}
\STATE{ Each agent $s$ executes action according to initial policy $\pi^b_{\mathrm{old}}(a^s_n|s'_{s,n})$ with $\boldsymbol{\theta}^b$.}
\STATE{ Obtain the reward $R_n$ and proceed to next state $s_{n+1}$}
\STATE{ Save ($\boldsymbol{s}_{d_{\mu}}(n)$, $\boldsymbol{a}_{d_{\mu}}(n)$, $r_{d_{\mu}}(n)$, $\boldsymbol{s}_{d_{\mu}}(n+1)$ for each agent.}
\ENDFOR
\STATE {Obtain a trajectory of each satellite $b: \{\boldsymbol{s}_{d_{\mu}}(n), \boldsymbol{a}_{d_{\mu}}(n), r_{d_{\mu}}(n)\}^N_{n=1}$}
\STATE {Calculate \{$\hat{Q}^b(s_n, \boldsymbol{a}_n) \}^N_{n=1}$ given in (\ref{Q_value})}	
\STATE {Calculate advantages $\{ A^b(s_n, \boldsymbol{a}_n) \}^N_{n=1}$ given in (\ref{advantage}) }		
\STATE {Store experience tuples $[\{s_{d_{\mu}}(n), a_{d_{\mu}}(n), \hat{Q}^b (s(n), \boldsymbol{a}(n)), A^b(s(n), \boldsymbol{a}(n))\}^B_{b=1}]^N_{n=1}$ into $\boldsymbol{L}$}
\FOR{$k=1,2,...,K$}
\STATE{ Shuffle and renumber the data’s order}
\FOR{j$=0,1,2,...,\frac{N}{O}-1$}
\STATE {Choose $B$ group of data $\boldsymbol{L}_j$:}
\STATE \indent {$\boldsymbol{L}_j = \big\{[s'_{s,i}, a_{s,i}, \hat{Q}^s_{s'_i, \boldsymbol{a}_i},  A^s(s'_i, \boldsymbol{a}_i)]^S_{s=1}\big\}^{B(J+1)}_{i=1+O_j}$} 
\FOR{$b=0,1,2,...,B$}
\STATE \indent {$\Delta \boldsymbol{\theta}^b = \frac{1}{O} \sum_{i=1}^{O} \Big\{ \nabla_{\boldsymbol{\theta}_b} \hat{\mathbb{E}} \big[ \mathrm{min}\{\Phi(\boldsymbol{\theta}_b)A_b(s(n),a(n)),$ \\ $\mathrm{clip}(\Phi(\boldsymbol{\theta}_b), \epsilon )A_b(s(n),a(n)) \}\big] \Big\}$}
\STATE \indent { $ \Delta \boldsymbol{\tau}^b = \frac{1}{O} \sum_{i=1}^{O} \bigg\{ \nabla_{\boldsymbol{\tau}_b} \hat{\mathbb{E}} \Big[ (\hat{Q}_b(s(n),a(n)) - Q_{\boldsymbol{\tau}^{\mathrm{old}}_b}(s(n),a(n)))^2\Big],   \bigg\} $}
\STATE \indent {Employ gradient ascent on $\boldsymbol{\theta}^b$ using $\Delta \boldsymbol{\theta}^b$ by Adam optimizer}
\STATE \indent {Employ gradient ascent on $\boldsymbol{\tau}^b$ using $\Delta \boldsymbol{\tau}^b$ by Adam optimizer}
\ENDFOR
\ENDFOR
\ENDFOR
\STATE { $\boldsymbol{\theta}^b_{\textrm{old}} \leftarrow \boldsymbol{\theta}^b$ and $\boldsymbol{\tau}^s_{\textrm{old}} \leftarrow \boldsymbol{\tau}^b$ for every satellite agent $b$ }
\STATE{Reset experience pool $\boldsymbol{L}$}
\ENDFOR
\STATE{\textbf{Output:} The optimal Co-MAPPO DRL network $\boldsymbol{\pi}_{\theta_{\textrm{opt}}}$}
\FOR{Each satellite $b=0,1,2,...,B$} 
\STATE{Calculate optimal bandwidth resource $\boldsymbol{Y}^*$ by closed-form solution by equation (\ref{KKT_sol_comm})} \COMMENT{In parallel on each satellite}
\STATE{Calculate optimal computing resource $\boldsymbol{\beta}^*$ by closed-form solution by equation (\ref{KKT_for_comp_1st}) and (\ref{KKT_for_comp_2nd})} \COMMENT{In parallel on each satellite}
\STATE{Calculate optimal CNS computing power ${\omega}^{h*}_{d_{\mu}}$ by closed-form solution by equation (\ref{Optimal_CNS})} \COMMENT{Only on CNS}
\ENDFOR
\end{algorithmic}
\end{algorithm}
\subsubsection{LMSs' and CubeSats computing Resource Allocation} The second subproblem concerning variable $\boldsymbol{\beta}$ is regarded with LMSs' and CubeSats' computing power allocation as below:
\begin{mini!}|s|[2]<b>
		{\substack{\boldsymbol{\beta}}}
		{\Upsilon_3 =  \sum_{\forall d \in \mathcal{D}} \sum_{\forall d_{\mu} \in d} \frac{\alpha_1 x^{b}_{d_{\mu}} T^{\mathrm{comp}}_{d_{\mu}, b} + \alpha_2 x^{b}_{d_{\mu}}P^{\mathrm{comp}}_{d_{\mu}, b} } {\|d\|\|D\|} } {\label{opt:P1.2.c}}{\textbf{P1.2.b:}}
		\addConstraint{\sum_{\forall d \in \mathcal{D}} \sum_{\forall \mu \in \mathcal{E}_b}  x^{b}_{d_{\mu}}\beta_{d_{\mu}} \leq \varrho_b, ~\forall b \in \mathcal{L} \cup \mathcal{C} {\label{C1.2.c}}}
\end{mini!}
The subproblem (\ref{opt:P1.2.c}) can be decomposed for each LMS and CubeSats $b$ as:
\begin{mini!}|s|[2]<b>
		{\substack{\boldsymbol{\beta}_b}}
		{\Upsilon^b_3 =  \sum_{\forall d_{\mu} \in d} \frac{\alpha_1 x^{b}_{d_{\mu}} T^{\mathrm{comp}}_{d_{\mu}, b} + \alpha_2 x^{b}_{d_{\mu}}P^{\mathrm{comp}}_{d_{\mu}, b} } {\|d\|\|D\|} } {\label{opt:P1.2.c1}}{}
		\addConstraint{\sum_{\forall d \in \mathcal{D}^b}  x^{b}_{d_{\mu}}\beta_{d_{\mu}} \leq \varrho_b.  {\label{C1.2.c1}}}
\end{mini!}
The variables $\boldsymbol{\beta}_{b} = {\beta_{d_{\mu}}},~\forall d_{\mu}\in \mathcal{D}^b$ are part of $\boldsymbol{\beta}$ which is linked with LMS or CubeSat $b$. By following KKT conditions as mentioned earlier, we find two solutions based on dual variable values, i.e., when $\iota_b = 0$ and $\iota_b \neq 0$ respectively:
\begin{equation} \label{KKT_for_comp_1st}
    \begin{aligned}
        \beta^*_{d_{\mu}} & = x^b_{d_{\mu}}\sqrt{\frac{\varphi}{\Lambda}},~\forall d_{\mu} \in \mathcal{D}^b, \\
         & \mathrm{where~~} \varphi_{d_{\mu}} = \frac{\alpha_1  \nu_d}{\eta_b  \|d\|  \|D\|},~\forall d_{\mu} \in \mathcal{D}^b,\\
         &  ~~~~~~~~~~~~ \Lambda_{d_{\mu}}  = \frac{\alpha_2 \chi_b  \eta_m}{\|d\|  \|D\|},~\forall d_{\mu} \in \mathcal{D}^b
    \end{aligned}
\end{equation}
\begin{equation} \label{KKT_for_comp_2nd}
    \begin{aligned}
        \beta^*_{d_{\mu}} & = x^b_{d_{\mu}}\sqrt{\frac{\varphi}{\Lambda + \iota_b}},\forall d_{\mu} \in \mathcal{D}^b,\\
        & \mathrm{where~~} \sum_{\forall d_{\mu} \in \mathcal{D}^b}  x^b_{d_{\mu}}\sqrt{\frac{\varphi}{\Lambda + \iota_b}} - \varrho_b = 0. 
    \end{aligned}
\end{equation}
\begin{figure}[t]
    \centering
    \includegraphics[width=0.9\columnwidth]{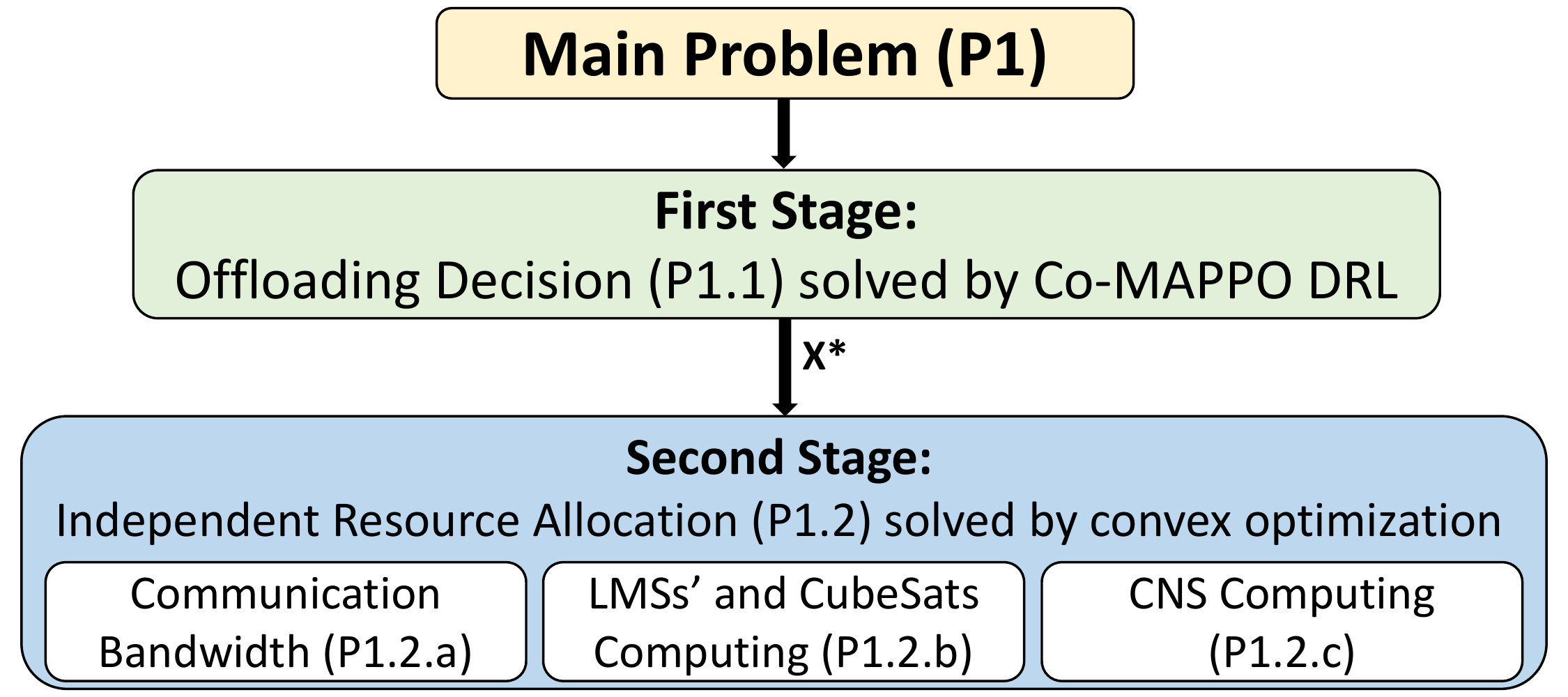}
    \caption{Illustration of solution flow chart and their information exchange.}
    \label{sol_flowchart}
    \vspace{-0.1in}
\end{figure}
The binary search can be used to find the value of $\iota_b$ in the second part of (\ref{KKT_for_comp_2nd}). For this subproblem, we have two closed-form solutions, i.e., (\ref{KKT_for_comp_1st}) and (\ref{KKT_for_comp_2nd}). However, we can select one that has a higher objective value. With the possession of local knowledge at each server $b$, we can deploy that resource allocation's solutions composed of (\ref{Optimal_CNS}), (\ref{KKT_sol_comm}), (\ref{KKT_for_comp_1st}), and (\ref{KKT_for_comp_2nd}).
\subsubsection{CNS Computing Resource Allocation} The third subproblem concerning $\boldsymbol{\omega}$ is regarded with CNS computing resource allocation, and it is written as below:
\begin{mini!}|s|[2]<b>
		{\substack{\boldsymbol{\omega}}^{h}_{d_{\mu}}}
		{\Upsilon_1 =  \alpha_1  \frac{\nu_{d_{\mu}}}{\omega^{h}_{d_{\mu}}} + \alpha_2  \chi_h  \omega^{c}_{d_{\mu}},} {\label{opt:P1.2.a}}{\textbf{P1.2.c:}}
		\addConstraint{ \omega^h_{d_{\mu}} \in R^+, ~\forall d_{\mu} \in d,~\forall d \in \mathcal{D}. {\label{C1.1.a.1}}}
\end{mini!}
The solution can be searched when the $\Upsilon_1$ gradient equates to zero, i.e., $\nabla \Upsilon_1=0$, and then solved for $\omega^h$. The derivation of $\nabla \Upsilon_1=0$ yields the optimal closed-form solution for computing resource allocation of each $d_{\mu}$ offloaded to the CNS $h$ as:
\begin{equation}
    \omega^{h*}_{d_{\mu}} = \sqrt{\frac{\alpha_1\nu_{d_{\mu}}}{\alpha_2 \chi_h}}, ~ \forall d_{\mu} \in d, \forall d \in \mathcal{D}.  \label{Optimal_CNS}
\end{equation}
In the following part, we will utilize these solutions for optimal resource allocation in a simulation environment. This work primarily concerns the challenge of jointly optimizing all satellites, given the coupling of association variables between ITS nodes and satellites. To address this issue, the proposed solution employs Co-MAPPO DRL in the initial stage to determine the value of offloading decision $\boldsymbol{X}$ variable. Once the offloading decision variable is established, each satellite, including LMSs and CubeSats, is responsible for its resource allocation, including communication and computation. Following this, the optimal offloading decision variable $\boldsymbol{X}^*$ is exchanged among all satellites, after which they independently make decisions regarding the remaining variables, as depicted in Figure \ref{sol_flowchart}. The complete algorithm for handling satellite-based ITS data is given in the Algorithm \ref{alg:MAPPO}.
\subsection{Proposed Algorithm Complexity}
The algorithm adopts a distributed implementation of the PPO algorithm, which enables each agent to optimize its policy by learning from both its own experiences and those of the other agents. The computational complexity of the Co-MAPPO algorithm is analyzed in terms of its training process and inference process. The time complexity of the training process is expressed as $O(T * N * n^2)$, where $T$ denotes the number of training iterations, $N$ is the number of agents, and $n$ is the number of parameters in the neural network. Moreover, the space complexity is given by $O(N * n^2 + M)$, where $M$ is the memory needed to store the experience pool. Regarding the inference process, the time complexity is $O(MD)$, where $M$ is the number of agents and $D$ is the dimensionality of the observation space. In addition to Co-MAPPO, this work examines the KKT conditions for solving linear programming problems, which provide a closed-form solution with polynomial time complexity. The KKT conditions involve a set of linear equations and inequalities, which can be efficiently solved by employing standard linear algebra methods. The overall worst-case time complexity of solving a linear system of equations using KKT conditions is expressed as $O(n^3)$, where $n$ is the number of variables in the linear system \cite{boyd2004convex}.
\setlength{\arrayrulewidth}{0.2mm}
\setlength{\tabcolsep}{1.4pt}
\renewcommand{\arraystretch}{1}
\begin{table}[t!]
\centering
\caption{Simulation Parameters}
\label{sim_tab}
\scalebox{1}{
\begin{tabular}{|l|l|}
\hline
    \textbf{Parameter}& \textbf{Value} \\ \hline \hline
    Random Required Task Memory & $[10, 90]~$~MB  \\ \hline
    Random Required Task computing Power & $[15, 70]~$~Gcycles/sec  \\ \hline
    CubeSats Bandwidth & $40~$~MHz  \\ \hline
    LMS Bandwidth & $200~$~MHz  \\ \hline
    CNS Bandwidth & $1~$~KHz  \\ \hline
    CubeSats computing Power & $10~$~Gigacycles/sec  \\ \hline
    LMS computing Power & $80~$~Gigacycles/sec \\ \hline
    CNS computing Power & $500~$~Gigacycles/sec \\ \hline
    Unit renting price of CubeSats computing  & $0.08~$\$/Gcycles \\ \hline
    Unit renting price of LMS computing  & $0.3~$\$/Gigacycles \\ \hline
    Unit renting price of CNS computing  & $10~$\$/Gigacycles \\ \hline
    Ka-band unit offloading price for CubeSats & $0.08\times10^{-4}~$\$/MB \\ \hline
    Ka-band unit offloading price for LMS & $0.12\times10^{-4}~$\$/MB \\ \hline
    Ka-band unit offloading price for CNS & $0.30\times10^{-4}~$\$/MB \\ \hline
    CTE's communication Power & $P_e~=~150~$mW \\ \hline
    Channel Gain & $G_{l,e}~=~5$dB \\ \hline
    Additive White Gaussian Noise & $N_0~=~10^{-5}$mW \\ \hline
    MST \& MSP weights & $\alpha_1,~\alpha_2~=0.5~$ \\ \hline
    Earth's radius $R_E$ & $6371~$km \\ \hline
    Earth's mass $M_E$ & $5.9722\times10^{24}~$kg \\ \hline
    Gravitational constant $G_E$ & $6.67\times10^{-11}$Nm$^2/$kg$^2$ \\ \hline
    Hidden layers \& Neurons in each layer & 3 , [512, 512, 512] \\ \hline
    Learning rate \& discount factor & 3e-4, 0.995 \\ \hline
    Experience memory \& batch size & 10240, 1024 \\ \hline
    Number of episodes & 1.5e+5 \\ \hline
\end{tabular}}
\end{table}
\begin{figure}[t]
    \centering
    \includegraphics[width=0.8\columnwidth]{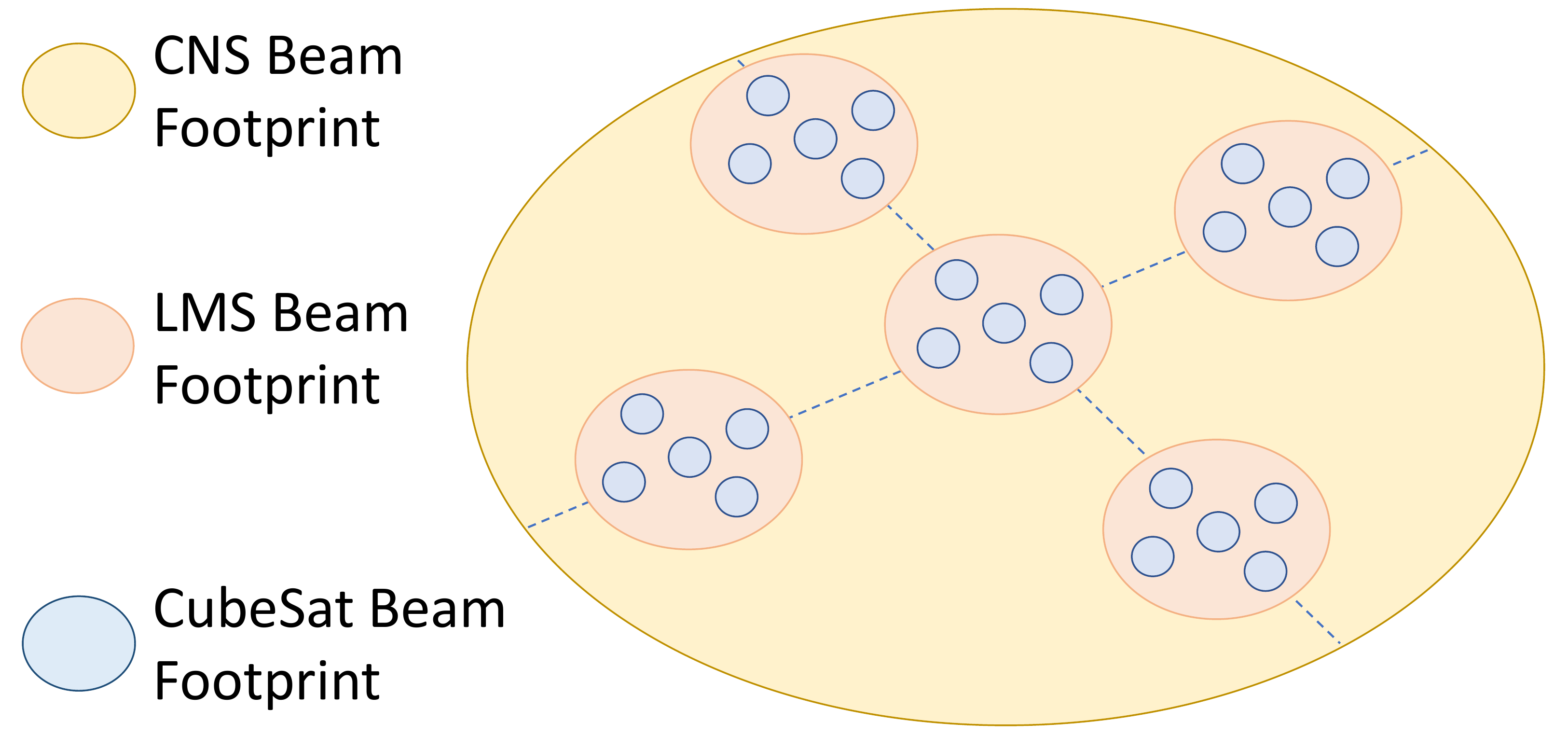}
    \caption{Illustration of each satellite footprint on the Earth.}
    \label{footprint}
    \vspace{-0.1in}
\end{figure}

\section{Performance Evaluation}
\label{sim_result}
\subsection{Simulation Settings}
For the simulation settings, we considered the $500~$km$^2$ region on earth where the beam footprint of CNS is constantly accessible. Due to the mobility of LMS and CubeSats, their services are available in periodic intervals over the considered region. A set of CubeSats and CTEs are defined as a neighboring set, which is under the coverage of each LMSs. Moreover, the CTEs have considered varying data memory depending upon their category. CTEs' arrival in a $2$D footprint follows a homogeneous Poisson point process (HPPP) distribution. The simulation settings are composed of satellite networks with $1~$CNS, $5~$LMS, and $25~$CubeSats. The CNS, LMS, and CubeSats are located at altitudes of $35786~$km, $1000~$km, and $200~$km, respectively, which need to serve $500$ CTEs. Each data-driven task is composed of five sub-tasks, suggesting that the accompanying data set is separated into five segments, each of which is held by an ITS node. Each ITS node is expected to have at least three CubeSats nearby for task offloading. The mobility of each satellite can be calculated by its circular orbit and altitude  \cite{cakaj2021parameters}. The radius of each satellite can be calculated with height $H^b$ as $R^b_{\mathrm{rad}}=H^b+R_{E}$ where $R_E$ is Earth's radius. The satellite's velocity can be defined as $V^b_{\mathrm{vel}} = \sqrt{\frac{M_E G_E}{R_{\mathrm{rad}}}}$, where $M$ Earth's mass and Gravitational constant. Thus, the orbital period for each satellite is $T^b_{\mathrm{orb}} = 2\pi\sqrt{\frac{R^3_{\mathrm{rad}}}{M_E G_E}}$. 
The data memory and computing resources of a task are selected from the intervals specified in Table \ref{sim_tab} along with the main parameters. Each satellite Earth beam footprint is shown in Fig. \ref{footprint}.
\subsection{Baselines}
To demonstrate the effectiveness of the sub-tasks allocation problem, we compare it with the following baselines:
\begin{itemize}
    \item CC-PPO: The central controller (CC) as a single-agent PPO to solve the problem (\ref{opt:P1}) with attention mechanism is referred to as CC-PPO. By considering the whole network as a super single agent, we specifically use PPO to derive centralized policy $\boldsymbol{\pi}$ \cite{WOARefBlue}.
    \item WOA: This algorithm utilizes the whale optimization algorithm (WOA) meta-heuristic solution approach. The WOA is used for meta-heuristic algorithm comparison due to its effectiveness, simplicity, and ability to solve complex optimization problems \cite{MIRJALILI201651RefBlue}.
    \item Random X: This algorithm considers that each LMS and CubeSat has equipped with a random offloading decision, which is trained with the local experience pool independently \cite{random_XRefBlue}.
\end{itemize}
\begin{figure*}[t]
        \begin{subfigure}[t]{0.33\textwidth}
                \includegraphics[width=\columnwidth, height=2.2in]{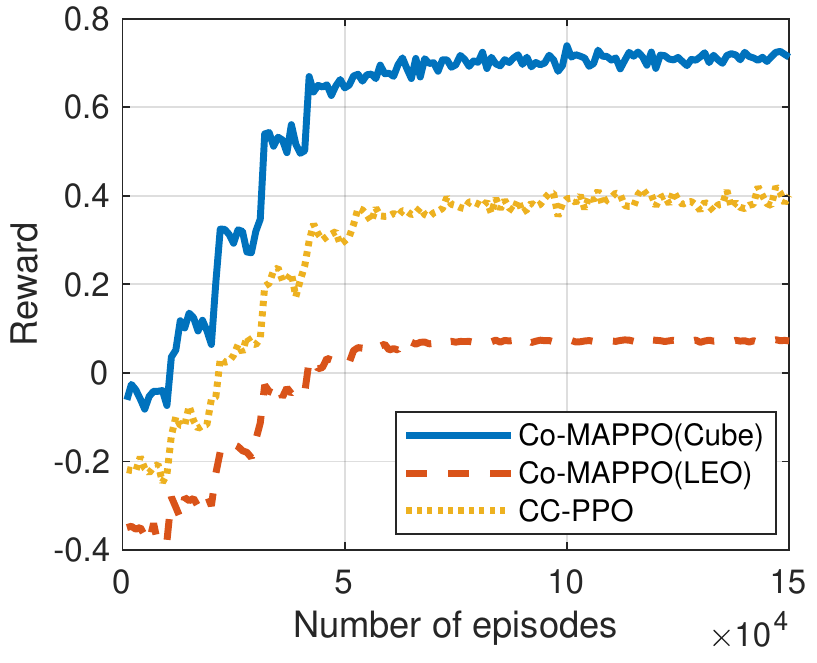}
                \caption{Reward convergence vs episodes for CubeSats, LMSs, and single agent ($\alpha_1, \alpha_2 = 0.5$).}
                \label{convergence_same_alphas}
        \end{subfigure}
        \begin{subfigure}[t]{0.33\textwidth}
                 \includegraphics[width=\columnwidth, height=2.2in]{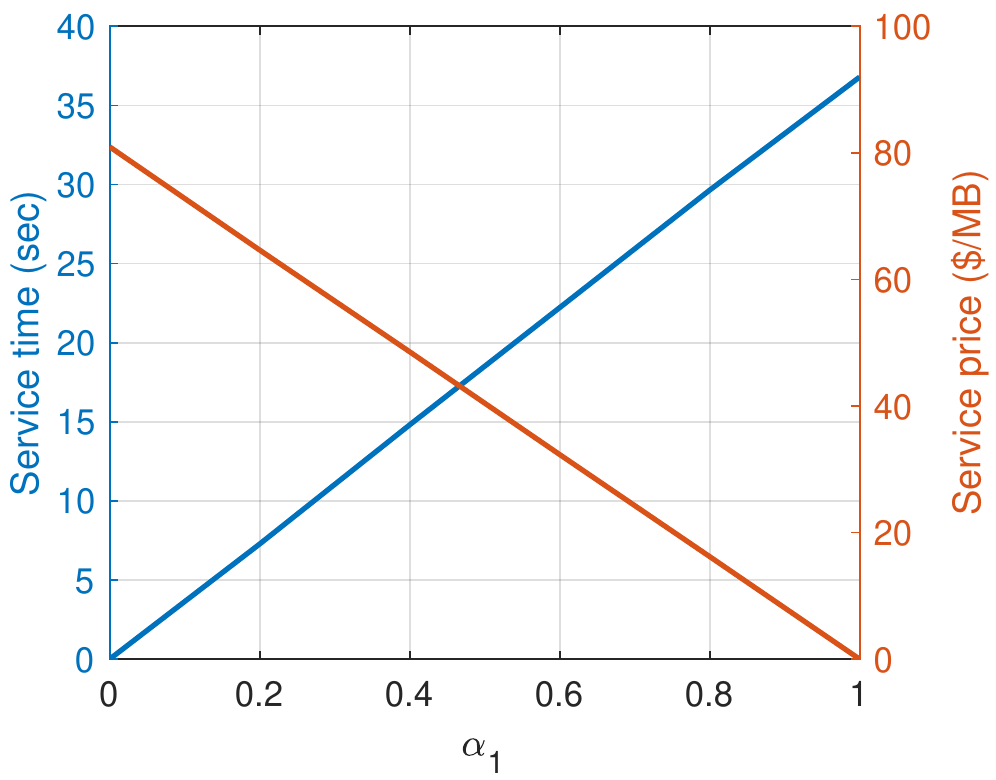}
                \caption{Trade-off between MST \& MSP vs $\alpha_1$.}
                \label{alpha_1}
        \end{subfigure}
        \begin{subfigure}[t]{0.33\textwidth}
                 \includegraphics[width=\columnwidth, height=2.2in]{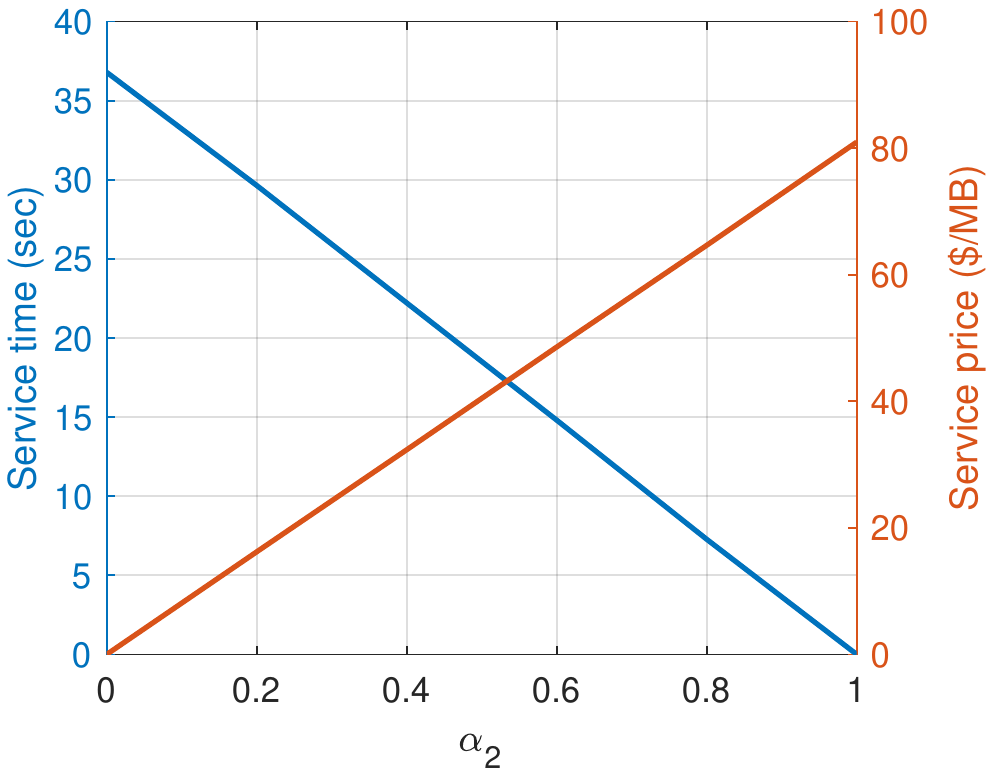}
                \caption{Trade-off between MST \& MSP vs $\alpha_2$.}
                \label{alpha_2}
        \end{subfigure}
        \caption{Illustration of the reward function convergence, and trade-off between $\alpha_1$ and $\alpha_2$ for MST and MSP, respectively.}
        \label{mixed1}
        \vspace{-0.2in}
\end{figure*}
\subsection{Experiment Results}
The convergence performance of the proposed method is depicted in Fig.~\ref{convergence_same_alphas}, where the change in reward is determined by the learning model episodes. We evaluated the proposed two agents' performance, i.e., CubeSats and LMSs, for sub-task selection and resource allocation using the proposed Co-MAPPO DRL algorithm. Then they were compared to CC-PPO, in which one agent executes all sub-task selection and resource allocation with no differentiation between agents. Each agent converged within a defined number of episodes. Following completion of the learning process, the CubeSat agent receives the largest reward, followed by the CC-PPO agent, and finally the LMS agent. These findings support our hypothesis that CubeSats are less costly and more appealing for computing than LMS. 

Figs.~\ref{alpha_1} and \ref{alpha_2} showcase the impact of weighting factors, namely  $\alpha_1$ and $\alpha_2$. The performance of Co-MAPPO DRL was analyzed to demonstrate the trade-off between MST and MSP when modifying  $\alpha_1$ and $\alpha_2$. A decrease in the value of  $\alpha_1$ corresponds to a reduction in the influence of MST, as its corresponding weight in the rewards is decreased, ultimately increasing MSP. Conversely, a decrease in the value of $\alpha_2$ corresponds to a reduction in the influence of MSP, as its corresponding weight in the rewards is decreased, ultimately increasing MST. Additionally, in a practical setting, the choice of $\alpha_1$ value is influenced by the degree of concern over latency. A lower  $\alpha_1$ value may be utilized if less latency is a priority, albeit at the cost of MSP; alternatively, a larger  $\alpha_1$ value may be utilized if the prioritization of MSP is paramount.

\begin{figure}[t]
    \centering
    \includegraphics[width=\columnwidth, height=3in]{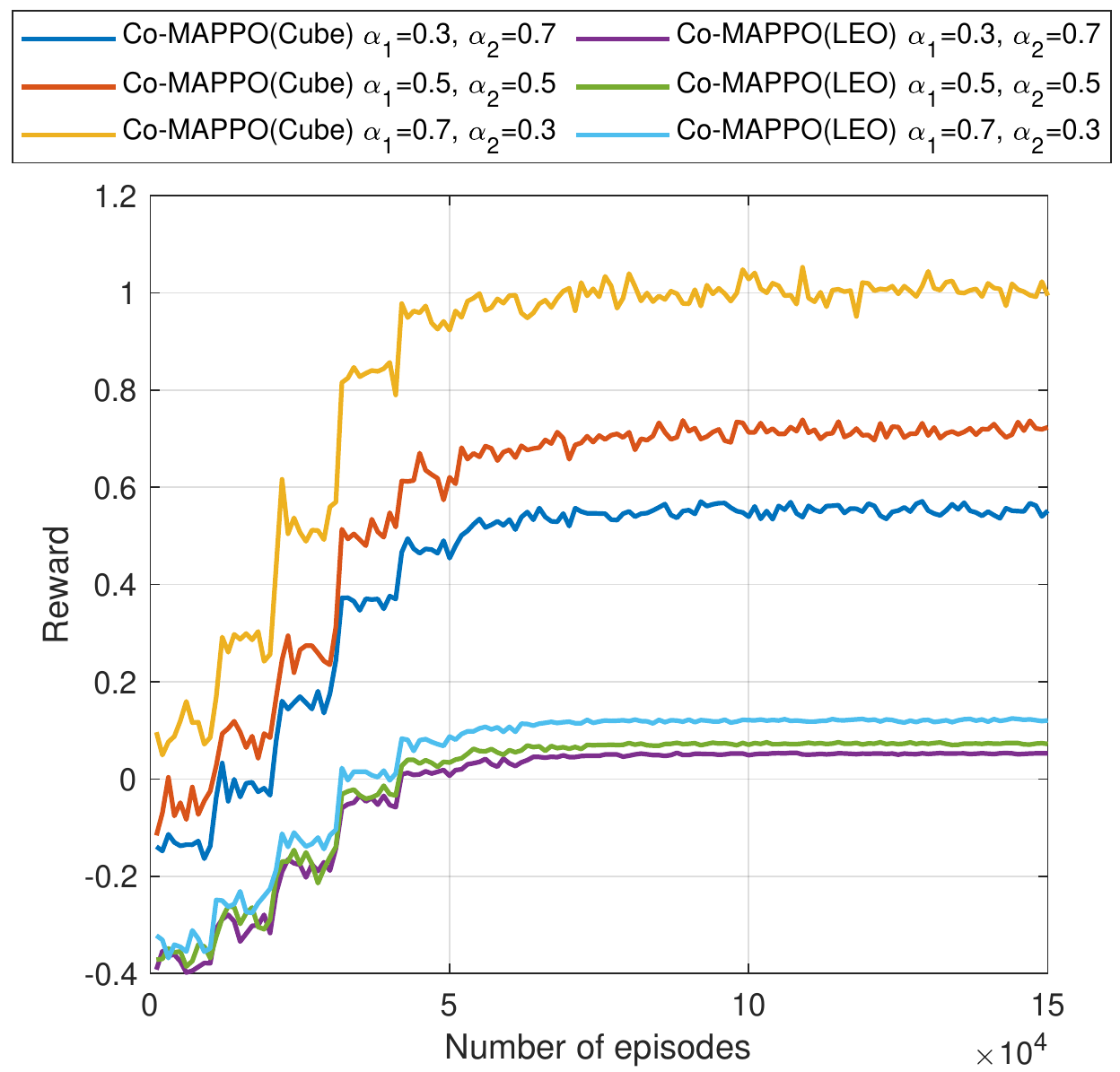}
    \caption{Cumulative rewards convergence according to $\alpha_1$ and $\alpha_2$ over the number of episodes for each agent.}
    \label{simulation_diff_alphas}
    \vspace{-0.1in}
\end{figure}

Fig.~\ref{simulation_diff_alphas} demonstrates the impact of the $\alpha_1$ and $\alpha_2$ weightage parameters on cumulative reward during the learning process when the proposed Co-MAPPO approach is employed. The $\alpha_1$ values were set to $0.3$, $0.5$, and $0.7$, while the $\alpha_2$ values were set to $0.7$, $0.5$, and $0.3$, respectively. The results show that an increase in $\alpha_1$ leads to an increase in final convergence compensation, indicating that the MST surpasses the absolute MSP. Furthermore, additional simulations were conducted with both parameters set at $0.5$. The same pattern was observed in both LMS and CubeSat agents. However, the cumulative rewards of CubeSat agents were higher than those of LMS agents due to their low prices and latency features.
\begin{figure*}[t]
    \centering
    \includegraphics[width=1.0\textwidth]{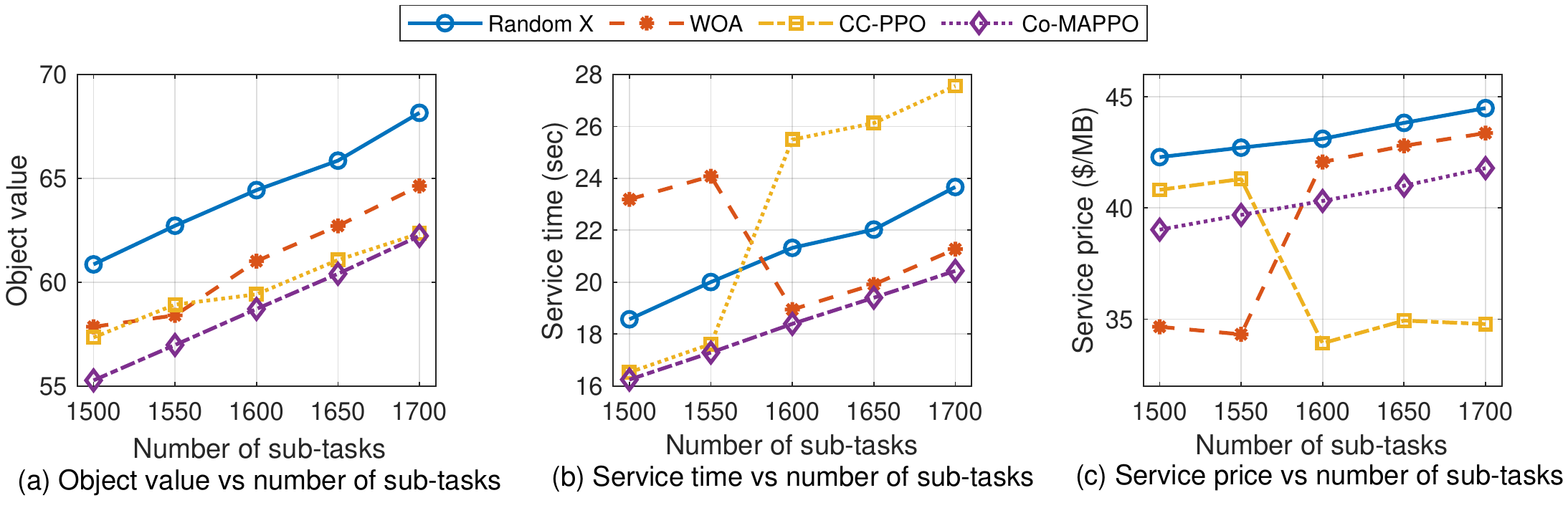}
    \caption{Comparison with benchmarks schemes for various number of sub-tasks.}
    \label{simulation_subtask_comparison}
\end{figure*}
\begin{figure*}[t]
    \centering
    \includegraphics[width=1.0\textwidth]{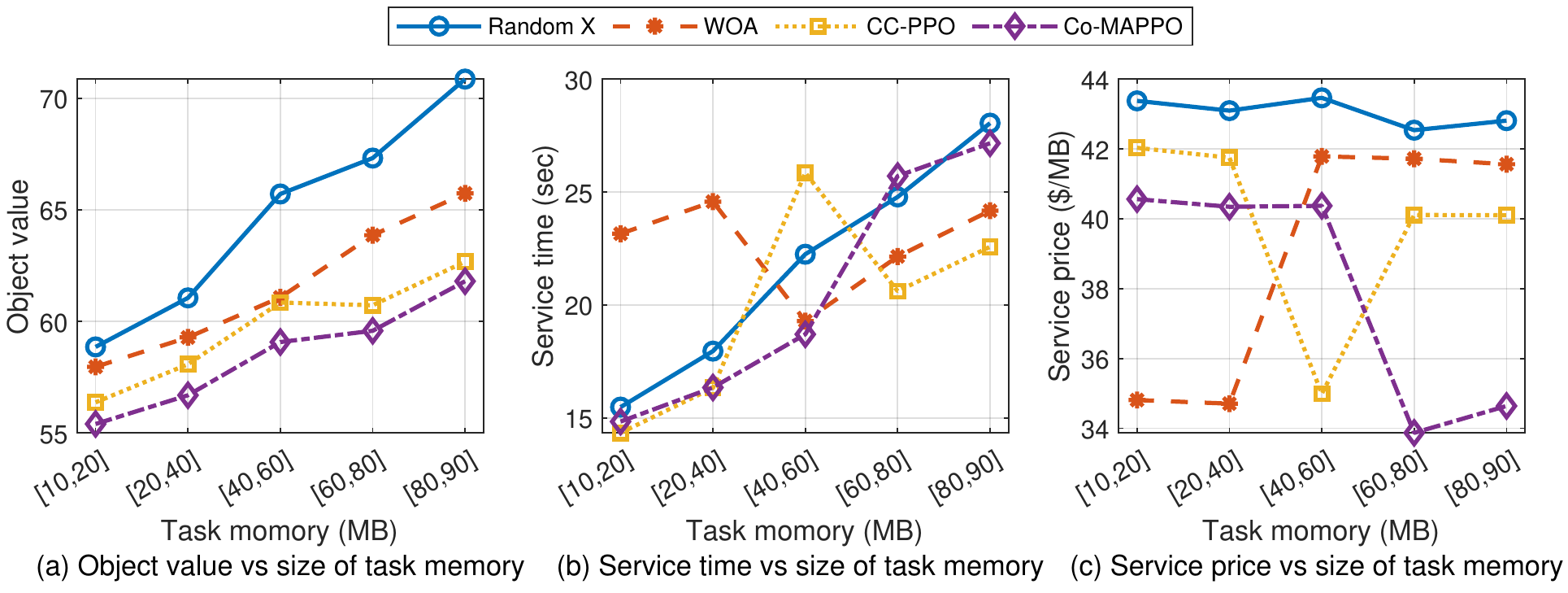}
    \caption{Comparison with benchmarks schemes for various task memory.}
    \label{simulation_taskmemory_comparison}
    \vspace{-0.1in}
\end{figure*}
\begin{figure*}[t]
    \centering
    \includegraphics[width=1.0\textwidth]{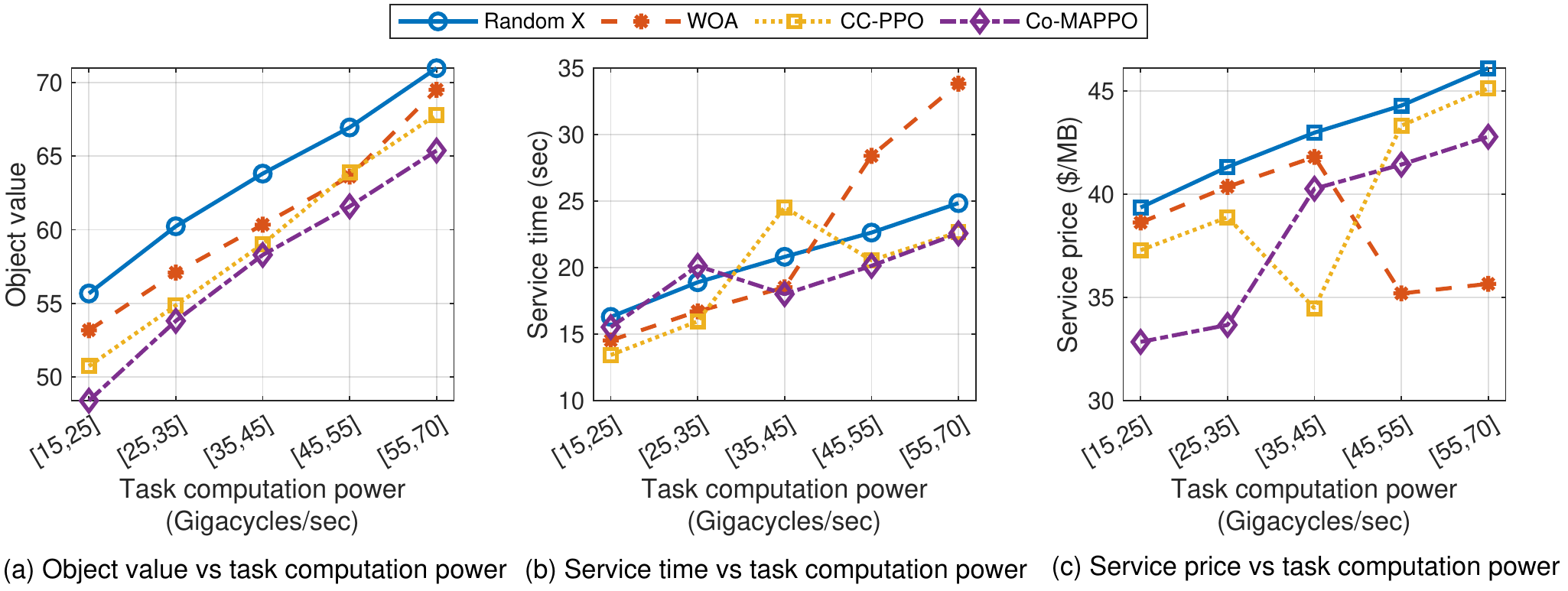}
    \caption{Comparison with benchmarks schemes for various task computation power.}
    \label{simulation_computationpower_comparison}
    \vspace{-0.1in}
\end{figure*}

We present a comparison of baselines for three main metrics: the overall objective function, MST, and MSP, concerning the number of sub-tasks. The results are illustrated in Fig.~\ref{simulation_subtask_comparison}. It is observed that the objective value in Fig.~\ref{simulation_subtask_comparison}a increases with the number of sub-tasks for all schemes, but the proposed approach (Co-MAPPO) obtains the lowest objective value, indicating that it outperforms the other baselines. Regarding MST, as shown in Fig.~\ref{simulation_subtask_comparison}b, the proposed scheme outperforms the baselines, whereas CC-PPO performs poorly due to the need for a higher number of communication rounds required for convergence. On the other hand, for MSP as shown in Fig.~\ref{simulation_subtask_comparison}c, the proposed schemes perform better initially when the number of sub-tasks is around $1550$, but as the number of sub-tasks increases, CC-PPO outperforms the proposed schemes, highlighting the trade-off between centralized and distributed agent settings in solving such problems. The superiority of CC-PPO, in theoretical terms, does not necessarily translate into its performance in practical implementations. It appears that the challenges of learning a centralized policy with a vast action space may account for its poor performance, as compared to learning multiple decentralized policies with a smaller action space. The Co-MAPPO scheme outperforms its counterparts, Random-X, WOA, and CC-PPO, in terms of the main objective function. Specifically, the Co-MAPPO scheme demonstrates a performance improvement of 9.9$\%$ when compared to Random-X, 5.2$\%$ when compared to WOA, and 4.2$\%$ when compared to CC-PPO with the number of sub-tasks is $1500$. These results suggest that the Co-MAPPO scheme holds promise for achieving superior outcomes in the proposed environment.

We also present a comparison of baselines for three metrics, namely, the overall objective function, MST, and MSP, concerning the size of task memory. The results of this analysis are depicted in Fig.~\ref{simulation_taskmemory_comparison}. The findings indicate that the objective value, as shown in Fig.~\ref{simulation_taskmemory_comparison}a, increases with the size of task memory for all schemes. However, the proposed approach, Co-MAPPO, achieves the lowest objective value, indicating superior performance compared to the other baselines. Regarding MST, the results in Fig.~\ref{simulation_taskmemory_comparison}b demonstrate that the proposed scheme performs better as the size of task memory increases. Additionally, the MST increases due to the limited computation resources of each satellite agent, and only the CNS option, with high latency, becomes available. Conversely, for MSP, as shown in Fig.~\ref{simulation_taskmemory_comparison}c, the proposed scheme does not initially perform better when the task size is around $40-60$. Nevertheless, as the size of task memory increases, the Co-MAPPO scheme outperforms all the baselines due to its superior optimization of the MST and MSP simultaneously for large task sizes. Further analysis indicates that the Co-MAPPO scheme outperforms its counterparts in terms of the main objective function. Specifically, the Co-MAPPO scheme demonstrates a performance improvement of $8.8\%$ compared to Random-X, $6.0\%$ compared to WOA, and $1.2\%$ compared to CC-PPO with a task memory size in the range of $80-90$~MB. These results suggest that the Co-MAPPO scheme holds promise for achieving superior outcomes in the proposed environment.
\begin{figure*}[t]
        \begin{subfigure}[t]{0.33\textwidth}
               \includegraphics[width=\columnwidth, height=1.9in]{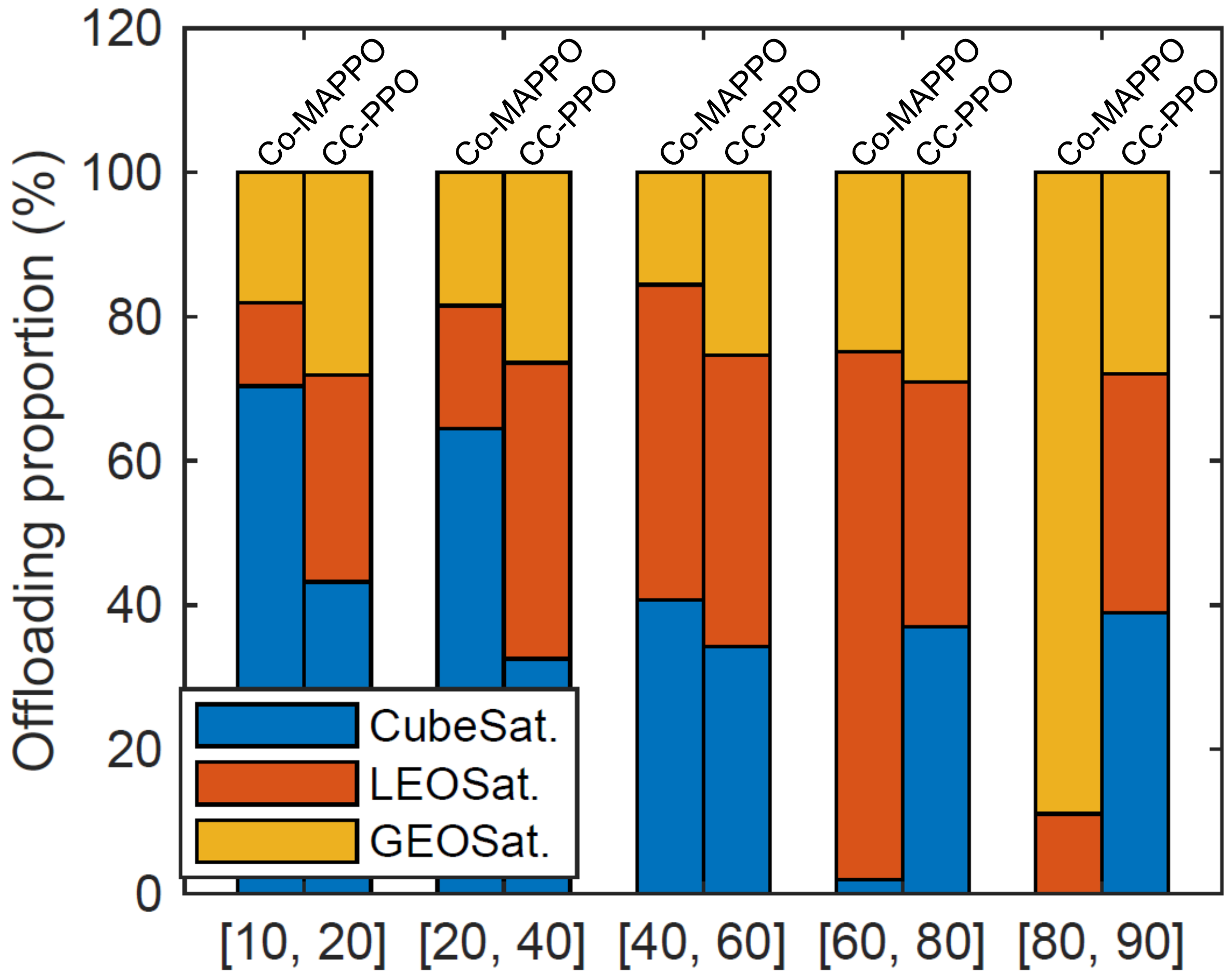}
                \caption{Sub-tasks offloading proportion according to the varying computing memory.}
                \label{memory}
        \end{subfigure}
        \begin{subfigure}[t]{0.33\textwidth}
               \includegraphics[width=\columnwidth, height=1.9in]{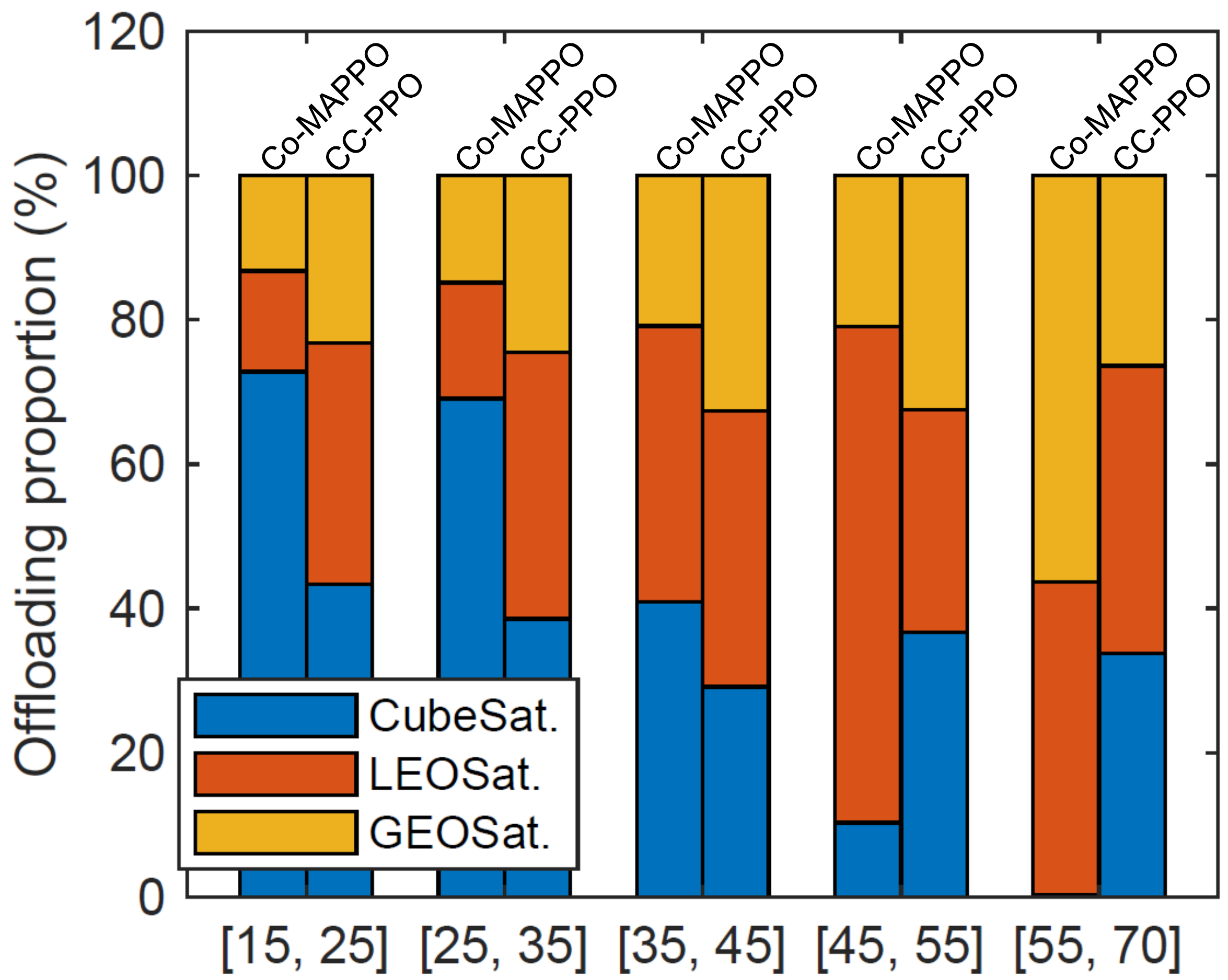}
                \caption{Sub-tasks offloading proportion according to the varying computing power.}
                \label{power}
        \end{subfigure}
        \begin{subfigure}[t]{0.33\textwidth}
                \includegraphics[width=\columnwidth, height=1.9in]{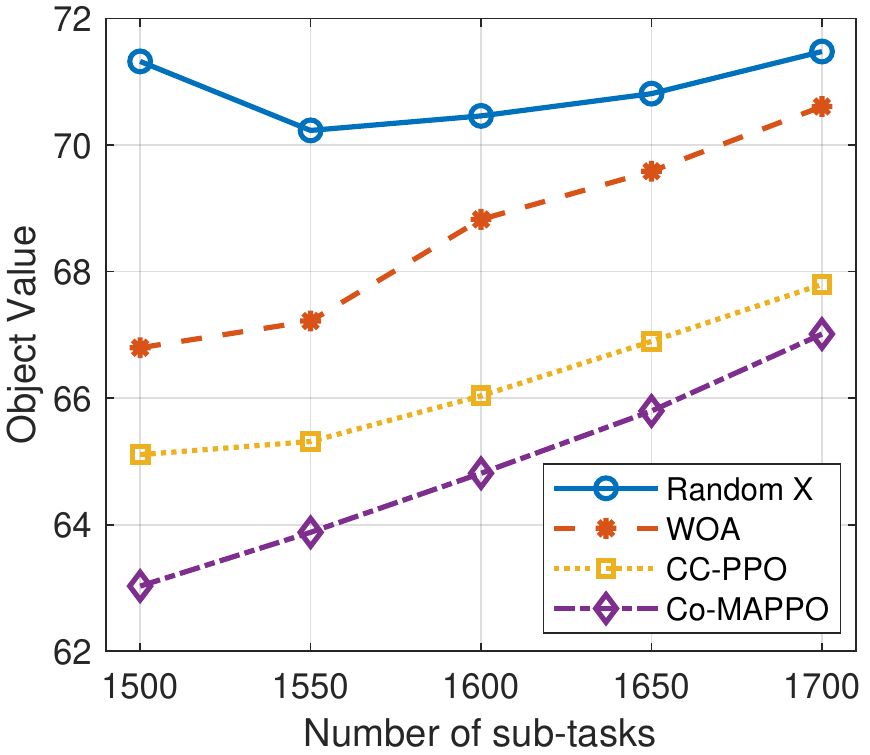}
                \caption{Comparison with non convex optimization for various sub-tasks (ablation study).}
                \label{simulation_woconvex_ablation}
        \end{subfigure}
        \caption{Illustration of proportional data offloading with various satellites and non-convex optimization comparison.}
        \label{mixed1}
        \vspace{-0.2in}
\end{figure*}

Similarly, we compare baselines for three metrics, i.e., overall objective function, MST, and MSP concerning the required task computation power. The results of this analysis are illustrated in Fig.~\ref{simulation_computationpower_comparison}. It was observed that the objective value, as shown in Fig.~\ref{simulation_computationpower_comparison}a, increases with the required task computation power for all schemes. However, the proposed approach, Co-MAPPO, achieved the lowest objective value, indicating superior performance compared to the other baselines. In terms of MST, the results depicted in Fig.~\ref{simulation_computationpower_comparison}b indicate that the proposed scheme did not perform better than the initial scheme when the required task computation power was lower. However, when the required task computation power increased, the system optimized the computational resources in a better way and outperformed the proposed schemes. This led to an increase in MST due to the better optimization of the computation resources of each satellite agent. In contrast, for MSP, as shown in Fig.~\ref{simulation_computationpower_comparison}c, the proposed scheme initially performed better when the required task computation power was around $25-35$. However, as the required task computation power increased, the Co-MAPPO scheme did not perform well, because of the higher prices needed to be spent to get more computation power from each agent. Further analysis indicates that the Co-MAPPO scheme outperforms its counterparts in terms of the main objective function. Specifically, the Co-MAPPO scheme demonstrated a performance improvement of $14.28\%$ compared to Random-X, $10.2\%$ compared to WOA, and $5.8\%$ compared to CC-PPO with a task memory size in the range of $15-25$ Gigacycle/sec. These results suggest that the Co-MAPPO scheme holds promise for achieving superior outcomes in the proposed environment.

Fig.~\ref{memory} demonstrate a comparison between the offloading task proportion status of each satellite and a range of computing memory requirements. The experimental study involved varying the required memory range between $10$ and $90$ megabytes for each sub-task. Our findings indicate that the proposed offloading approach allocated proportionately to CNSs as the required memory increased while allocating to CubeSat when the memory requirement was low. Conversely, the allocation through CC-PPO demonstrated no significant correlation, indicating a low-performance level. These results highlight the significance of the proposed Co-MAPPO DRL approach over the CC-PPO method. Additionally, Fig.~\ref{power} illustrates the comparison of the offloading proportion status of each satellite with a range of required computing power. The experimental study involved generating the required power for each sub-task within the range of 15 to 70 Gigacycles per second. The results demonstrate a similar trend as observed in Fig.~\ref{memory}, where the proposed approach allocated proportionately to CNSs as the required power increased and to CubeSat when the power requirement was low. These findings further validate the proposed offloading approach.

Fig. \ref{simulation_woconvex_ablation} illustrates the results of the ablation study conducted to investigate the performance of the objective function for a proposed scheme in comparison to benchmark schemes. The Co-MAPPO DRL algorithm is utilized without incorporating the convex optimization component, and each decision variable is learned through the training procedure. The proposed algorithm exhibits superior performance as compared to the benchmark schemes, indicating an improvement of $11.8\%$ when compared to Random-X, $5.9\%$when compared to WOA, and $3.0\%$ when compared to CC-PPO, with the same number of sub-tasks (i.e., $1500$). Furthermore, when compared to the configuration involving convex optimization, the proposed Co-MAPPO DRL algorithm yields a performance improvement of $14.5\%$, owing to its ability to allocate resources based on an offloading decision that enhances the overall performance. However, a trade-off exists between the two configurations: while learning the resource allocation variable necessitates more computational resources, an analytical solution enables the system to make an instant decision using a closed-form solution.

\section{Conclusion}
\label{conc}
The present study examines a novel service scenario of data-driven ITS task offloading within MEC-enabled multi-layer satellite networks. In this context, heterogeneous computing servers, i.e., CNS, LMSs, and CubeSats, are expected to collaborate to process sub-tasks offloaded by various ITS nodes. The task offloading problem is defined by specifying the features of data-driven tasks, including communication and computing methods of diverse satellites, as well as the associated rental price to concurrently decrease MST and MSP. Initially, the Co-MAPPO DRL with an attention algorithm is developed to decide task offloading in a distributed manner. Subsequently, the problem of resource allocation is divided into three subproblems based on the convex theory, and an optimal closed-form analytic solution is obtained for each problem using KKT conditions. Finally, a simulation model is constructed, and comprehensive simulation results illustrate the superiority of the proposed approach when compared with baselines. In future work, the integration of terrestrial base stations with semantic communication among network nodes will be explored.





\bibliographystyle{IEEEtran}
\bibliography{ref}
\let\mybibitem\bibitem

\begin{IEEEbiography}[{\includegraphics[width=1in,height=1.25in, clip,keepaspectratio]{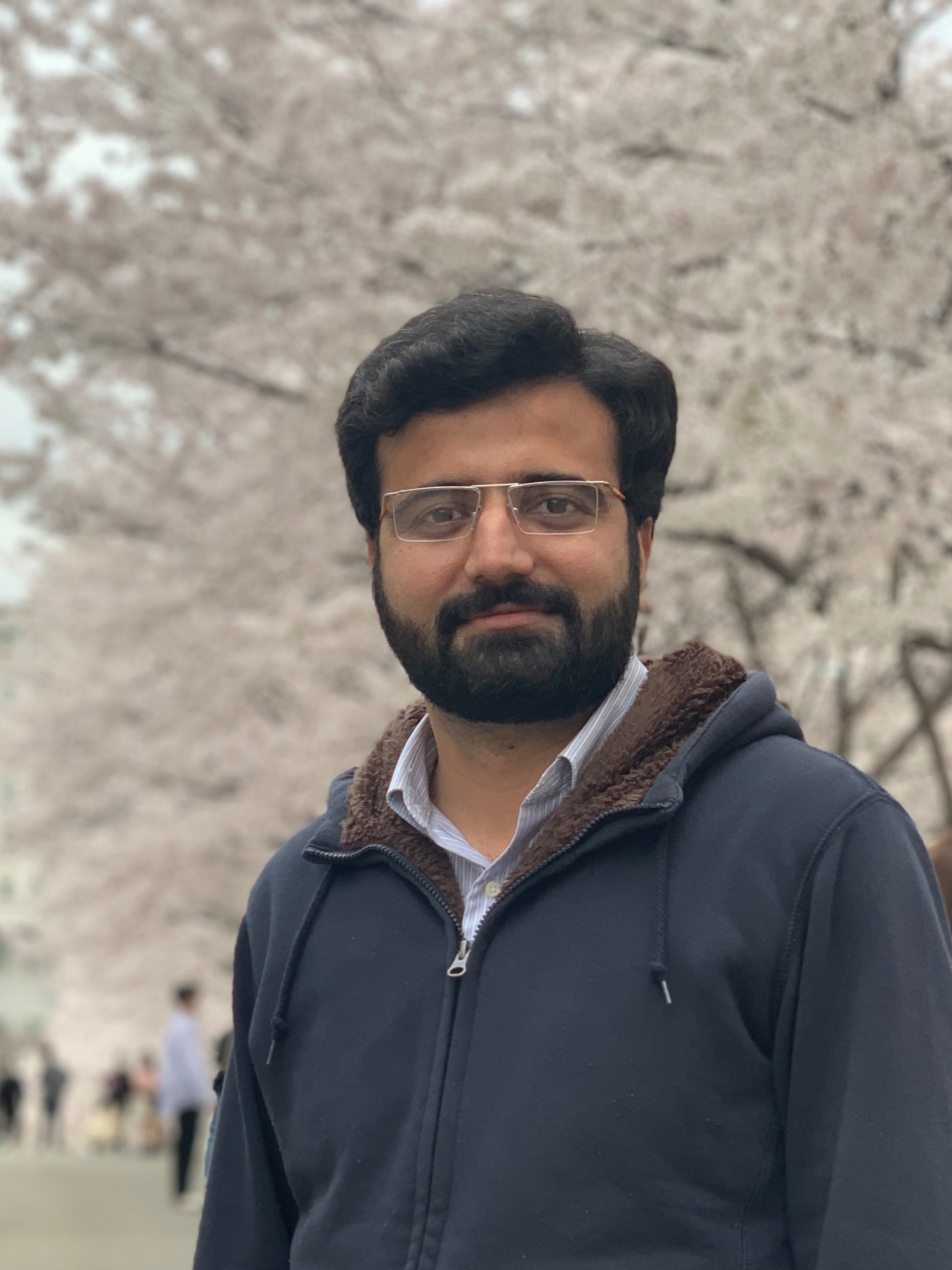}}]{Sheikh Salman Hassan}(S'14) received his BS (Electrical Engineering) degree magna cum laude from the National University of Computer and Emerging Sciences (NUCES-FAST), Karachi, Pakistan in 2017. He is currently pursuing a Ph.D. (Computer Science \& Engineering) degree at Kyung Hee University (KHU), Republic of Korea. He received the Best Poster and Paper Award at the International Conference on Information Networking (ICOIN) 2021 and 2023, respectively. His research interests include 6G, non-terrestrial networks, the Internet of Everything, and intelligent network management.
\end{IEEEbiography}
\begin{IEEEbiography}[{\includegraphics[width=1in,height=1.25in, clip,keepaspectratio]{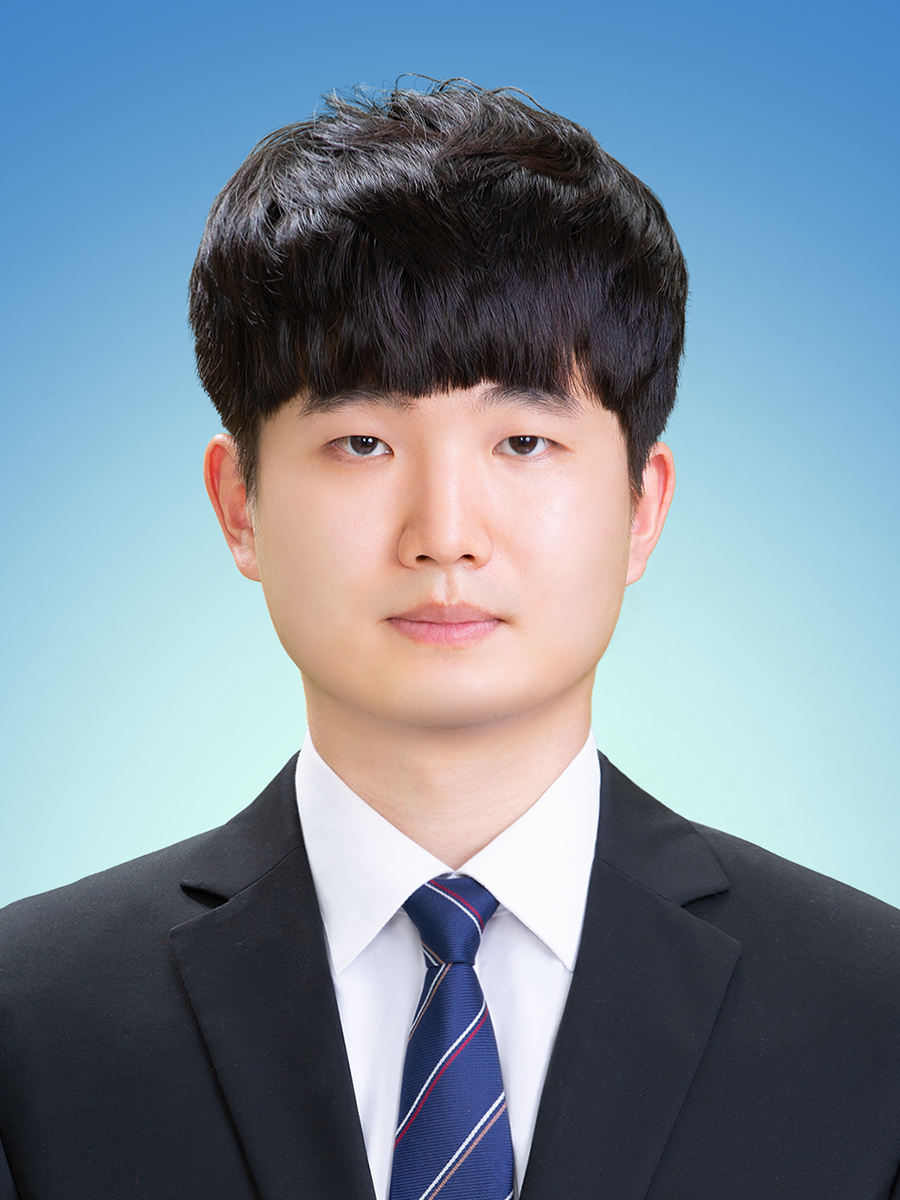}}]{Yu Min Park} received the B.S. degree in Applied Mathematics and Computer Engineering from Kyung Hee University, South Korea, in 2019, and the M.S. degree in Computer Engineering from Kyung Hee University, South Korea, in 2021. He is currently pursuing a Ph.D. degree in Computer Engineering, at Kyung Hee University, South Korea. His research interests include reinforcement learning, intelligent networking management system, and network resource optimization.
\end{IEEEbiography}
\begin{IEEEbiography}[{\includegraphics[width=1in,height=1.25in, clip,keepaspectratio]{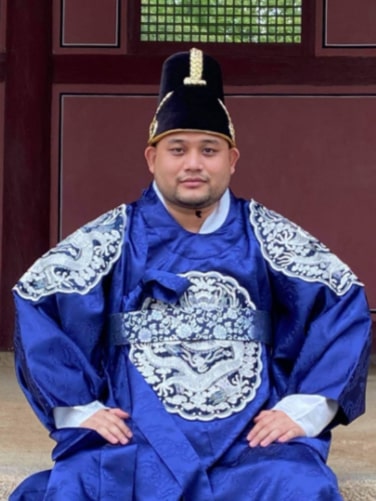}}]{Yan Kyaw Tun} (M’ 22) received the B.E. degree in marine electrical systems and electronics engineering from Myanmar Maritime University, Thanlyin, Myanmar, in 2014 and the Ph.D. degree in Computer Science and Engineering from Kyung Hee University, Seoul, South Korea, in 2021. He worked as a postdoc in the Intelligent Networking Lab. Currently, he is working as a postdoc in the School of Electrical Engineering and Computer Science, KTH Royal Institute of Technology, Stockholm, Sweden. He received the best Ph.D. thesis award in Engineering in 2020. His research interests include network economics, game theory, network optimization, wireless communication, wireless network virtualization, mobile edge computing, and wireless resource slicing for 5G.
\end{IEEEbiography}
\begin{IEEEbiography}[{\includegraphics[width=1in,height=1.25in,clip,keepaspectratio]{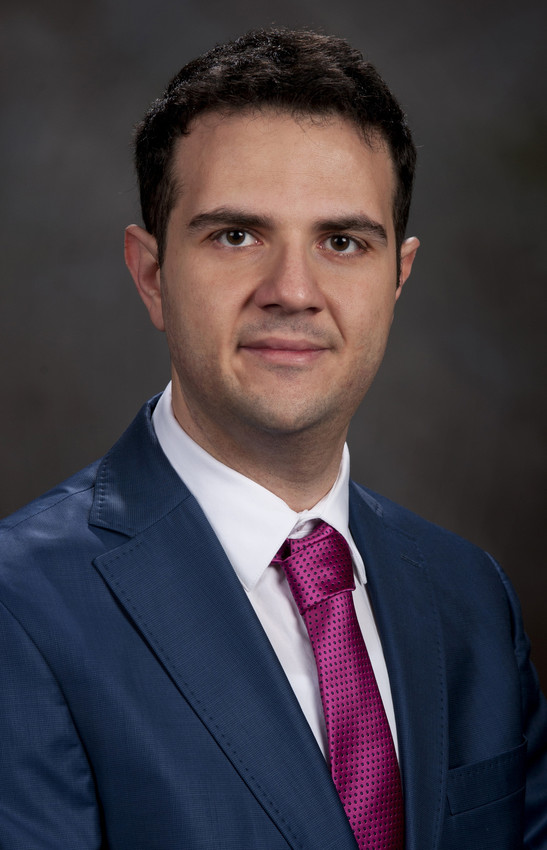}}]{Walid Saad} (S’07, M’10, SM’15, F’19) received his Ph.D degree from the University of Oslo in 2010. He is currently a Professor at the Department of Electrical and Computer Engineering at Virginia Tech,
where he leads the Network sciEnce, Wireless, and Security (NEWS) laboratory. His research interests include wireless networks (5G/6G/beyond), machine learning, game theory, security, unmanned aerial vehicles, semantic communications) cyber-physical systems, and network science. Dr. Saad is a Fellow of the IEEE. He is also the recipient of the NSF CAREER award in 2013, the AFand OSR summer faculty fellowship in 2014, and the Young Investigator Award from the Office of Naval Research (ONR) in 2015. He was the author/co-author of eleven conferences best paper awards at WiOpt in 2009, ICIMP in 2010, IEEE WCNC in 2012, IEEE PIMRC in 2015, IEEE SmartGridComm in 2015, EuCNC in 2017, IEEE GLOBECOM in 2018, IFIP NTMS in 2019, IEEE ICC in 2020 and 2022, and IEEE GLOBECOM in 2020. He is the recipient of the 2015 and 2022 Fred W. Ellersick Prize from the IEEE Communications Society, of the 2017 IEEE ComSoc Best Young Professional in Academia award, of the 2018 IEEE ComSoc Radio Communications Committee Early Achievement Award, and of the 2019 IEEE ComSoc Communication Theory Technical Committee. He was also a co-author of the 2019 IEEE Communications Society Young Author Best Paper and of the 2021 IEEE Communications Society Young Author Best Paper. From 2015-2017, Dr. Saad was named the Stephen O. Lane Junior Faculty Fellow at Virginia Tech and, in 2017, he was named College of Engineering Faculty Fellow. He received the Dean’s award for
Research Excellence from Virginia Tech in 2019. He was also an IEEE Distinguished Lecturer in 2019- 2020. He currently serves as an editor for the IEEE Transactions on Mobile Computing and the IEEE Transactions on Cognitive Communications and Networking. He is an Area Editor for the IEEE Transactions on Network Science and Engineering, an Associate Editor-in-Chief for the IEEE Journal on Selected Areas in Communications (JSAC) Special issue on Machine Learning for Communication Networks, and an Editor-at-Large for the IEEE Transactions on Communications. He is the Editor-in-Chief for the IEEE Transactions on Machine Learning in Communications and Networking.
\end{IEEEbiography}
\begin{IEEEbiography}[{\includegraphics[width=1in,height=1.25in,clip,keepaspectratio]{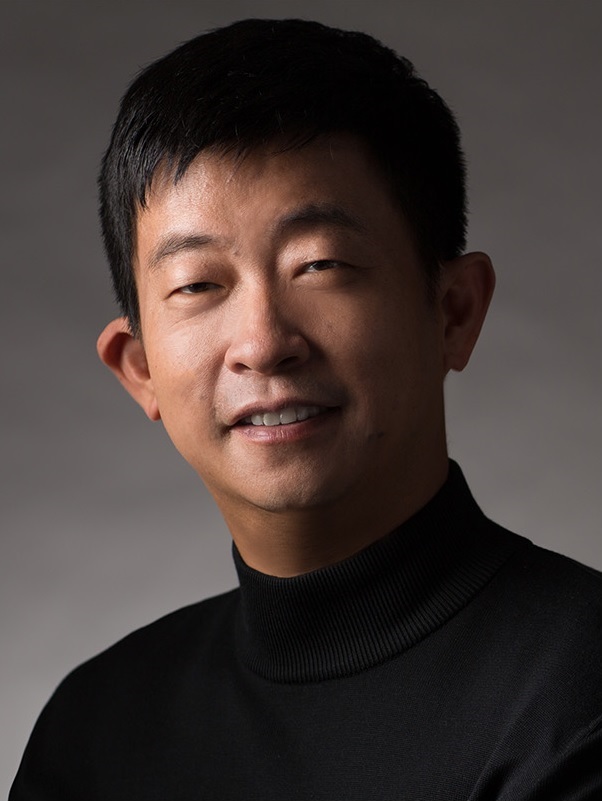}}]{Zhu Han}(S’01, M’04, SM’09, F’14) received the B.S. degree in electronic engineering from Tsinghua University, in 1997, and the M.S. and Ph.D. degrees in electrical and computer engineering from the University of Maryland, College Park, in 1999 and 2003, respectively. From 2000 to 2002, he was an R\&D Engineer of JDSU, Germantown, Maryland. From 2003 to 2006, he was a Research Associate at the University of Maryland. From 2006 to 2008, he was an assistant professor at Boise State University, Idaho. Currently, he is a John and Rebecca Moores Professor in the Electrical and Computer Engineering Department as well as in the Computer Science Department at the University of Houston, Texas. He is also a Chair professor in National Chiao Tung University, ROC. His research interests include wireless resource allocation and management, wireless communications and networking, game theory, big data analysis, security, and smart grid.  Dr. Han received an NSF Career Award in 2010, the Fred W. Ellersick Prize of the IEEE Communication Society in 2011, the EURASIP Best Paper Award for the Journal on Advances in Signal Processing in 2015, IEEE Leonard G. Abraham Prize in the field of Communications Systems (best paper award in IEEE JSAC) in 2016, and several best paper awards in IEEE conferences. Dr. Han was an IEEE Communications Society Distinguished Lecturer from 2015-2018, and is AAAS fellow since 2019 and ACM distinguished Member since 2019. Dr. Han is 1\% highly cited researcher since 2017 according to Web of Science.\end{IEEEbiography}
\begin{IEEEbiography}[{\includegraphics[width=1in,height=1.25in, clip,keepaspectratio]{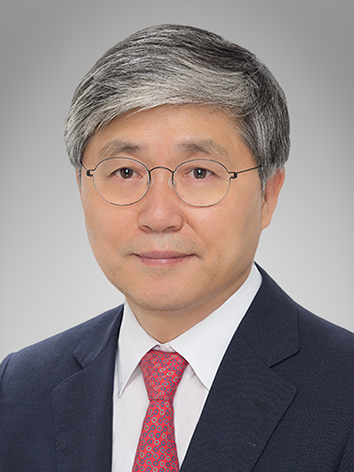}}]{Choong Seon Hong} (S’95-M’97-SM’11)  received the B.S. and M.S. degrees in electronic engineering from Kyung Hee University, Seoul, South Korea, in 1983 and 1985, respectively, and the Ph.D. degree from Keio University, Tokyo, Japan, in 1997. In 1988, he joined KT, Gyeonggi-do, South Korea, where he was involved in broadband networks as a member of the Technical Staff. Since 1993, he has been with Keio University. He was with the Telecommunications Network Laboratory, KT, as a Senior Member of Technical Staff and as the Director of the Networking Research Team until 1999. Since 1999, he has been a Professor with the Department of Computer Science and Engineering, Kyung Hee University. His research interests include future Internet, intelligent edge computing, network management, and network security.  Dr. Hong is a member of the Association for Computing Machinery (ACM), the Institute of Electronics, Information and Communication Engineers (IEICE), the Information Processing Society of Japan (IPSJ), the Korean Institute of Information Scientists and Engineers (KIISE), the Korean Institute of Communications and Information Sciences (KICS), the Korean Information Processing Society (KIPS), and the Open Standards and ICT Association (OSIA). He has served as the General Chair, the TPC Chair/Member, or an Organizing Committee Member of international conferences, such as the Network Operations and Management Symposium (NOMS), International Symposium on Integrated Network Management (IM), Asia-Pacific Network Operations and Management Symposium (APNOMS), End-to-End Monitoring Techniques and Services (E2EMON), IEEE Consumer Communications and Networking Conference (CCNC), Assurance in Distributed Systems and Networks (ADSN), International Conference on Parallel Processing (ICPP), Data Integration and Mining (DIM), World Conference on Information Security Applications (WISA), Broadband Convergence Network (BcN), Telecommunication Information Networking Architecture (TINA), International Symposium on Applications and the Internet (SAINT), and International Conference on Information Networking (ICOIN). He was an Associate Editor of the IEEE TRANSACTIONS ON NETWORK AND SERVICE MANAGEMENT and the IEEE JOURNAL OF COMMUNICATIONS AND NETWORKS. He currently serves as an Associate Editor for the International Journal of Network Management.
\end{IEEEbiography}

\end{document}